\documentclass{emulateapj}

\newcommand{\msun}{M$_{\odot}$}
\newcommand{\zsun}{Z$_{\odot}$}

\shorttitle{Star Clusters in NGC\,922}
\shortauthors{Pellerin et al.}

\usepackage{natbib}
\begin{document}


\title{The Star Cluster Population of the Collisional Ring Galaxy NGC 922}


\author{Anne~Pellerin\altaffilmark{1}}
\affil{George P. and Cynthia Woods Mitchell Institute for Fundamental Physics and Astronomy, \\
Department of Physics and Astronomy, Texas A\&M University, 4242 TAMU, College Station, TX 77843}
\email{pellerin@physics.tamu.edu} 

\author{Gerhardt~R. Meurer\altaffilmark{2}}
\affil{Physics \& Astronomy Department, Johns Hopkins University, 3400 Charles Street, Baltimore, MD 21218, USA}
\email{Gerhardt.Meurer@icrar.org}

\author{Kenji Bekki\altaffilmark{2}}
\affil{Department of Astrophysics \& Optics, School of Physics, University of New South Wales, NSW, 2052, Australia} {\email{bekki@phys.unsw.edu.au} 

\author{Debra~M. Elmegreen}
\affil{Vassar College, Department of Physics and Astronomy, Box 745, Poughkeepsie, NY 12604, USA}
\email{elmegreen@vassar.edu}

\author{O.~Ivy Wong}
\affil{Astronomy Department, Yale University, P.O. Box 208101, New Haven, CT 06520-8101, USA}
\email{ivy.wong@yale.edu}
\and 

\author{Patricia~M. Knezek}
\affil{WIYN Consortium Inc., PO Box 26732, Tucson, AZ 85726, USA}
\email{knezek@noao.edu}

\altaffiltext{1}{Former address: Physics \& Astronomy Department, Johns Hopkins University, 3400 Charles Street, Baltimore, MD 21218, USA}
\altaffiltext{2}{Present address: ICRAR / University of Western Australia, 35 Stirling Highway, Crawley, WA 6009, Australia}

\begin{abstract}
We present a detailed study of the star cluster population detected in the galaxy NGC\,922, one of the closest collisional ring galaxies known to date, using HST/WFPC2 UBVI photometry, population synthesis models, and N-body/SPH simulations.
We find that 69\% of the clusters are younger than 7\,Myr, and that most of them are located in the ring or along the bar, consistent with the strong H$\alpha$ emission. The cluster luminosity function slope of 2.1-2.3 for NGC\,922 is in agreement with those of young clusters in nearby galaxies. 
Models of the cluster age distribution match the observations best when cluster disruption is considered. We also find clusters with ages ($>$50\,Myr) and masses ($>$10$^5$\,\msun) that are excellent progenitors for faint fuzzy clusters. 
The images also show a tidal plume pointing toward the companion. Its stellar age from our analysis is consistent with pre-existing stars that were stripped off during the passage of the companion.
Finally, a comparison of the star-forming complexes observed in NGC\,922 with those of a distant ring galaxy from the GOODS field indicates very similar masses and sizes, suggesting similar origins.
\end{abstract}


\section{Introduction}
\label{intro}

Collisional ring galaxies represent a very specific case of galaxy interaction for which a companion drops through a much more massive spiral galaxy \citep[e.g.][]{lynds76,theys77,hern93}. First it requires very specific initial conditions to occur. The mass of the companion must be small compared to the main galaxy and the primary must be a disk galaxy. Second, the characteristic ring of star formation only lasts for a few 10$^8$\,yr, which is not a large temporal window compared to the Hubble timescale. Although they are rare objects in the local universe, they are thought to be more common at high redshifts \citep{lav96,elm06}. 

NGC\,922 is a drop-through ring galaxy for which the perturber, responsible for the ring pattern, has recently been discovered \citep[][hereafter W06]{wong06}. Using simulations, the authors found that NGC\,922's companion (named S2) interacted gravitationally with NGC\,922 about 330\,Myr ago through an off-centered collision, creating the C-shape morphology seen in the disk of NGC\,922. The dwarf compact companion is about 20 times less massive than NGC\,922, the latter having an estimated mass of 2.8$\times$10$^{10}$\,\msun. NGC\,922 is one of the nearest collisional ring galaxies known, at a distance of 43\,Mpc \citep{meurer06}, about three times closer than the iconic Cartwheel Galaxy \citep[e.g.][]{fos77,dav82,str96}. It has a heliocentric velocity of 3077\,km\,s$^{-1}$, and an estimated star formation rate (SFR) of 7-8\,\msun\,yr$^{-1}$. A more detailed description of the physical properties of both NGC\,922 and S2 can be found in the work of W06. 

%% FIGURE 1 %%
\begin{figure*}[!ht]
\epsscale{1.15}
%\plottwo{ngc922_uvi_980pc_CMYK.eps}{J0224-24_trim.eps}
\plottwo{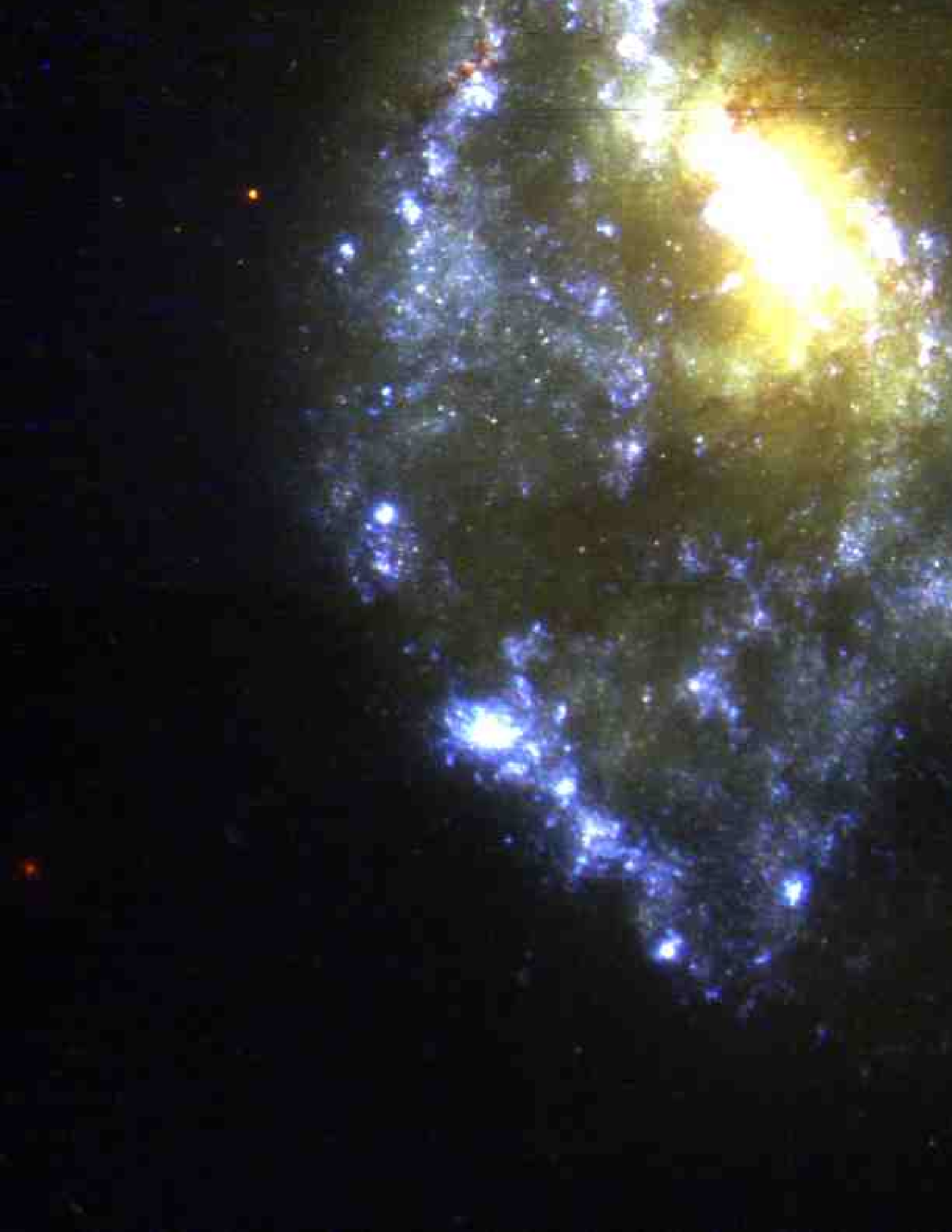}{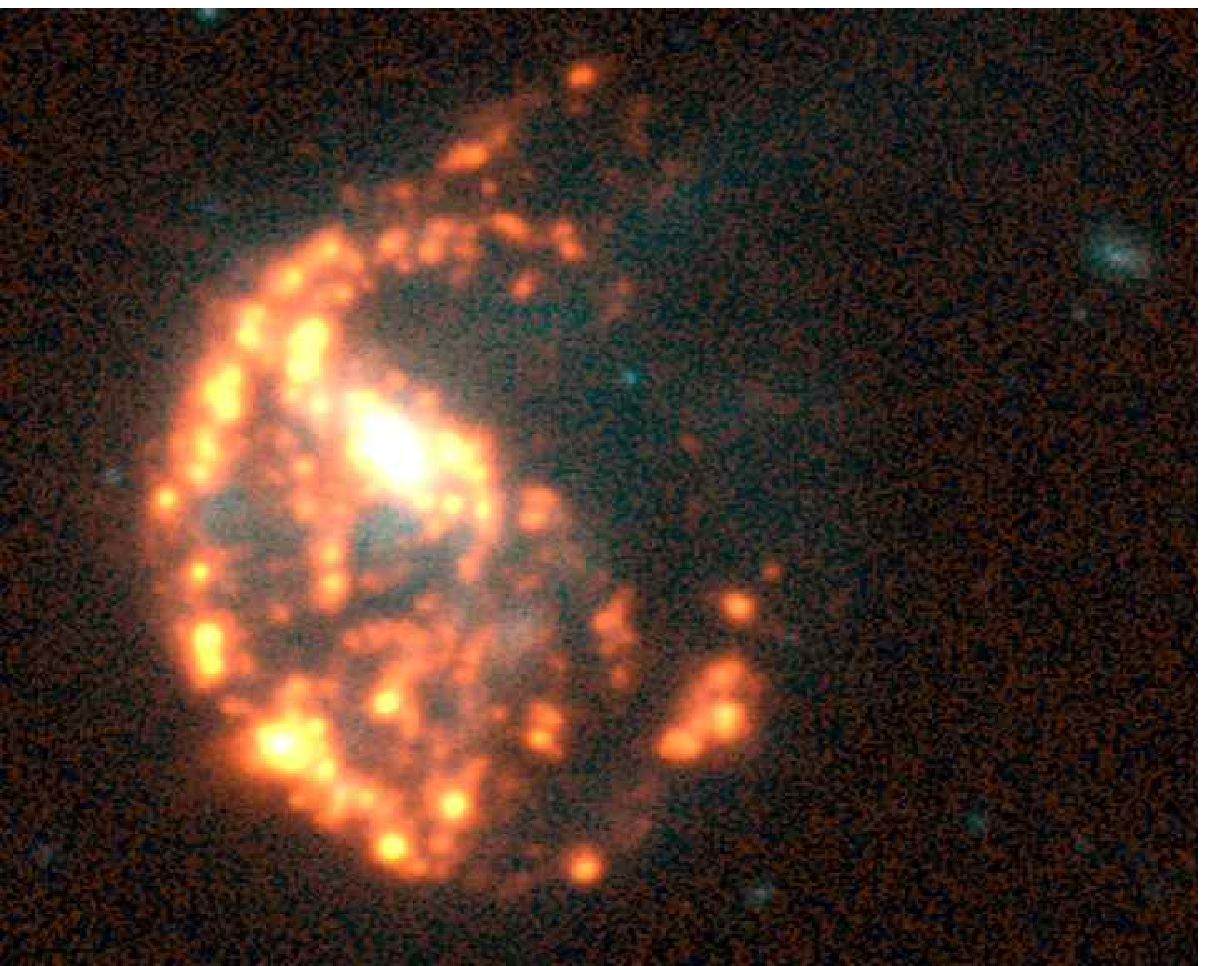}
\caption{\label{cmyk} Left: Three-color HST/WFPC2 image of NGC\,922. Blue: F300W; Green: F547M; Red: F814W. The image is 2.5$\arcmin\times$2.0$\arcmin$ ($\sim$30.7$\times$24.6\,kpc$^2$). Right: H$\alpha$ image of NGC922 from SINGG. The resolution is $\sim$1.5$\arcsec$, lower than the HST image. Blue: R-band; Green: H$\alpha$+ continuum; Red: H$\alpha$ with continuum substracted. North is up and east is left.\\}
\end{figure*}

\section{Data processing and PSF Photometry}
\label{data}

NGC\,922 is a very interesting object to study. Since the companion responsible for the star formation activity is known, dynamical simulations can be better constrained and its recent star formation history can be modeled. Furthermore, because of its proximity, its stellar cluster population can be resolved and observed with the {\it Hubble Space Telescope} (HST). In this paper, we present recent HST optical images of NGC\,922. The overall goal is to study the recent star cluster formation history of NGC\,922 and to compare it to the dynamical models in order to constrain the models and the recent dynamical history of NGC\,922 as well as better understanding the star formation and cluster evolution processes.

In the next section, we present the HST data as well as the photometric analysis of the star clusters detected in NGC\,922. In section~\ref{syn} we present the synthesis modeling performed to estimate cluster ages and masses. In section~\ref{pop} we describe the star cluster properties. Section~\ref{test} discusses the effect of cluster dissolution on the observed star cluster population. We then present an N-body/SPH simulation of the collision and compare the results to the observations (\S\ref{model}). In section~\ref{large} we investigate the large-scale star formation of NGC\,922 and compare it to distant ring galaxies. Section~\ref{tail} presents diffuse features detected in NGC\,922. We then discuss the fate of NGC\,922 and its star clusters in section~\ref{fate}. We conclude in \S\ref{conclusion}.

Six sets of images of NGC\,922, two slightly different pointings with three exposures each, have been obtained on 25 and 26 June 2007 using the Wide Field Planetary Camera 2 (WFPC2) onboard HST in the four filters F300W, F439W, F547M, and F814W (hereafter U, B, V, and I for simplicity). The F439W and F547M filters were chosen over F450W and F555W to avoid contamination by the nebular emission lines of [O\,{\sc{iii}}] and H$\beta$ in the passbands. The WFPC2 field covers the whole ring galaxy, but excludes S2.
%, which is located approximately 100\,kpc North-West of NGC\,922's nucleus. 

The data were first reduced by the STScI calibration pipeline, which includes treatment for the bias, dark, and flat fields. Then we corrected for warm pixels, geometric distortion and flux calibration, as well as charge transfer efficiency. We used the variable-pixel linear reconstruction method (or drizzling) and the IRAF task {\it {multidrizzle}} to combine the individual exposures, remove cosmic rays and to improve the spatial resolution of the final images. The chips were drizzled individually to an output pixel size of 0.075$\arcsec$ pixel$^{-1}$ (15.6\,pc~pixel$^{-1}$) for the WF chip. %This leads to a spatial resolution of 39\,pc. 

A three-color image from our HST/WFPC2 data is shown in Figure~\ref{cmyk} together with a H$\alpha$ image from the Survey for Ionization in Neutral-Gas Galaxies \citep[SINGG;][]{meurer06}.
%(SINGG\footnote{http://www.stsci.edu/ftp/science/singg/}). 
Figure~\ref{morph} presents the main morphological features of NGC\,922 over-plotted on the F814W WFPC2 image. The blue ring has a projected diameter of 15\,kpc and is particularly prominent on the eastern side, giving a ``C" like morphology to NGC\,922. While the eastern side of the ring displays sharp edges with star-forming regions well distributed along it, the western side shows a patchy distribution and weak diffuse H$\alpha$ emission. A 9\,kpc off-center bar, which appears yellow in Fig.~\ref{cmyk}a, displays spokes emanating from it. A relatively isolated star-forming region is also observed on the South-West portion of the ring but is disconnected from the ``C". Another star-forming structure, like an accent over the ``C", is observed north of the ring, and merges quite well with the ring near the bar end. On the H$\alpha$ image an inner arc can be seen internal to the ring, perhaps as an extension of the ``accent". However it is impossible to conclude here if this structure is physically real or just a random distribution of star-forming regions misleading the eye.

%% FIGURE 2 %%
\begin{figure*}
\epsscale{0.75}
%\plotone{ngc922_f814w_morph.eps}
\plotone{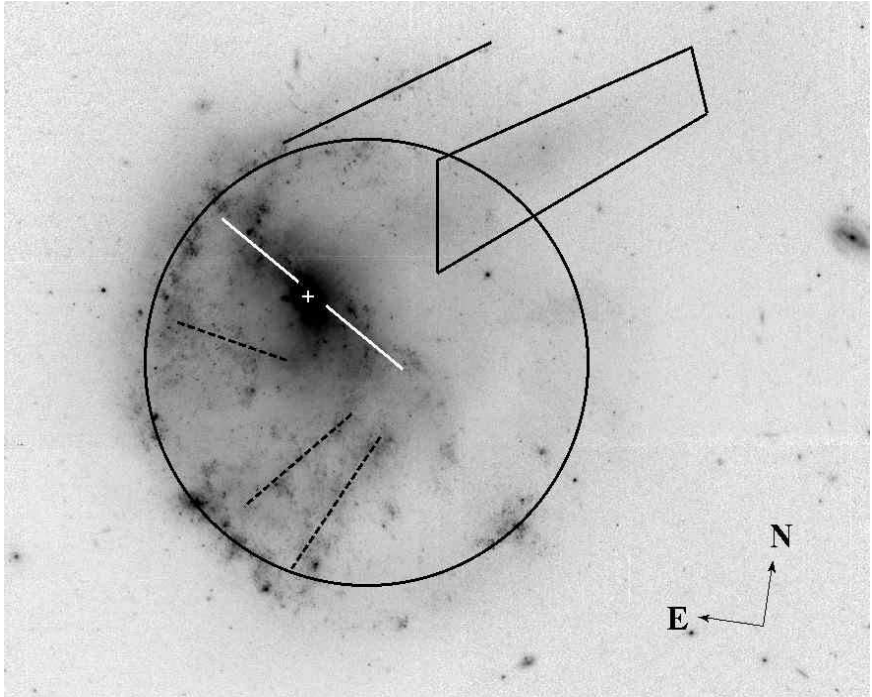}
\caption{\label{morph} Morphological features of NGC\,922 over-plotted on the F814W image. The circle shows the expanding ring of star formation. The $+$ symbol identifies the nucleus. The white lines follow the bar. The dotted lines indicate the location of possible spokes. The polygon highlights the tidal plume. The solid line, North of the circle, shows a feature of star formation outside of the expanding ring. The physical scale is the same as in Fig.~\ref{cmyk}.\\}
\end{figure*}

To study the star cluster properties, we performed PSF photometry using the IRAF/DAOPHOT package. PSF photometry is required because of crowding. For each chip and each pointing, we used bright unsaturated sources to create a PSF model of 4.5 pixels radius and a FWHM$\sim$2.5 pixels, which slightly varies depending on the chip. In the case of the PC1 chip, there were too few sources to create a reliable PSF model. For this reason, and because the same region was observed on the WF2 chip of the other pointing, we are using only the photometry based on the WF chips for this work. A summary of the observational datasets is presented in Table~\ref{hstobs}. The absolute photometric calibration is in the Vegamag system and was obtained with the zero points calculated for each chip as described on the STScI/WFPC2 cookbook website\footnote{http://www.stsci.edu/hst/wfpc2/analysis/wfpc2\_cookbook.html}. Because the Multidrizzle software converts the WFPC2 images from DN to e- in order to properly reject cosmic rays, we adjusted the photometric zeropoint to account for this gain. We calculated and applied the aperture correction for each of the four filters and chips.

%TABLE 1
\begin{deluxetable*}{lccccc}
\tabletypesize{\scriptsize}
\tablecaption{\label{hstobs} HST/WFPC2 Observational Data Parameters of NGC\,922}
\tablehead{
\colhead{Filter} & \colhead{Pointing} & \colhead{Exp. Time} & \colhead{Detection Limit} & \colhead{Chip} & \colhead{PSF FWHM} \\
\colhead{} & & sec & mag &  & pixels
}
\startdata
F300W & A &1100 & 23.8 & WF2 & 2.5 \\
            & A &          &         & WF3 & 3.3 \\ 
            & A &          &         & WF4 & 2.6 \\
            & B & 1100 &         & WF2 & 2.5 \\
            & B &           &         & WF3 & 2.8 \\
F439W & A &600 & 24.3 & WF2 & 2.3 \\
            &  A &       &        & WF3 & 2.5 \\
            &  A &       &        & WF4 & 2.3 \\
            & B & 600 &        & WF2 & 2.3 \\
            & B &        &        & WF3 & 2.5 \\
F547M & A &600 & 26.2 & WF2 & 2.3 \\
            & A &        &        & WF3 & 2.5 \\
            & A &        &        & WF4 & 2.5 \\
            & B & 600 &        & WF2 & 2.4 \\
            & B &        &        & WF3 & 2.1 \\
F814W & A &600 & 27.1 & WF2 & 2.2 \\
            & A &        &        & WF3 & 2.8 \\
            & A &        &        & WF4 & 2.3 \\
            & B &600  &        & WF2 & 2.5 \\
            & B &        &        & WF3 & 2.5 \\
\enddata 
\end{deluxetable*}

To determine the photometric uncertainties and the completeness level of each cluster, we generated a list of artificial sources of various magnitudes and background levels for each combination of chip and filter. This is critical to take into account that more clusters will remain undetected in regions of high level background (often linked to crowding), and that the photometry will also be less accurate. We used this list as input for the DAOPHOT/addstar task to generate aritificial sources in the WFPC2 images. We then performed the PSF photometry again on the new images with added stars. We used the results to create functions of photometric uncertainties and completeness levels that takes into account the background level. Then we used these functions to associate the proper photometric uncertainty and completeness level values to each detected source and each filter.  On average, the 50\% completeness limits are reached at 23.8, 24.3, 26.2, and 27.1 magnitudes for the U, B, V, and I filters respectively. We also applied a foreground extinction correction of E(B$-$V)=0.02 \citep{sch98}. We find that these detection limits may include individual bright stars among faint cluster samples (see discussion in \S\ref{distrib}). Finally we do not expect much contamination from background galaxies, i.e. about 5 to 15 galaxies over the entire field, to our exposure depth, based on the Hubble Deep Field \citep{sha98}.

We created four lists for all the clusters detected in each of the four filters. We then cross-correlated the four lists to find sources with matching coordinates. The matching objects had to be within 2.5~pixels from each other to be considered the same physical source. Overall we observed 2248 star clusters\footnote{Due to the very large number of clusters, the detailed photometry is not published here, but is available upon request.} detected in the four filters simultaneously through PSF photometry (hereafter UBVI sample). An additional 1569 clusters were also observed in the three BVI filters (hereafter BVI sample), but displayed no conterpart in the U-band image. Because of the given spatial resolution of 39\,pc (2.5~pixels), it is probable that some of the sources detected contain multiple clusters. A completeness correction was applied to each cluster found in each filter and based on the completeness level determined in that filter. For simultaneous detections in multiple filters, we combined the completeness corrections of the appropriate filters.

An examination of the residual images showed that three extended sources were significantly bigger than the PSF size, and were therefore badly fitted. They comprise two large clusters and the nucleus. For those sources, we performed a simple aperture photometry using polygons and the IRAF task {\it {polyphot}}. The photometric results for those three sources are given in Table~\ref{complex}.

\section{Stellar Population Modeling}
\label{syn}

We used the population synthesis model Starburst99\footnote{http://www.stsci.edu/science/starburst99/} \citep{lei99} to estimate cluster ages and masses. We used an instantaneous burst of star formation with a standard initial mass function \citep[IMF;][]{kroupa02} ranging from 0.1 to 100\,\msun, as well as the Padova stellar evolutionary tracks with stars from the asymptotic giant branch. A solar metallicity was chosen to match the result of W06 and based on the nuclear spectrum from the 6dF Galaxy Redshift Survey \citep{jones04}. For comparison, we also performed the synthesis using different model properties with 1) an IMF ranging from 0.1 to 30\,\msun\ and 2) with a metallicity of 0.4\,\zsun. However, considering the photometric uncertainties, these two additional models did not give significantly different results.

For an accurate comparison between the stellar population models and our photometric data, we applied the IRAF/SYNPHOT package on the Starburst99 spectra to calculate the theoretical magnitudes that would be seen in our four WFPC2 filters. 
%A distance modulus of 33.2 mag was applied to the theoretical photometry to fit NGC\,922's distance. 
Then we performed a $\chi^2$ fitting on the photometric data points, basically the slope of the spectral energy distribution (SED), to obtain the best-fit age. Finally, we calculated the best-fit mass for each cluster by averaging the magnitude differences between the observation and the models in the four filters for the UBVI sample, and in the three filters for the BVI sample. The mass and age uncertainties were estimated by fitting the models within the photometric error bars using the bootstrap technique.
%and using down to three photometric points only.

Most of the clusters found in this galaxy are very young, with a mean age of 16\,Myr and a median age of 6\,Myr for the UBVI sample, and a mean age of 34\,Myr and a median age of 6\,Myr for the BVI sample. Interestingly, the sources of the BVI sample are mainly faint sources and we obtained young ages for most of them. 
%This means that the U-band non-detection is mainly due to either a low mass cluster or a high extinction rather than an older cluster age. 

\subsection{Photometric Modeling and Extinction}

%TABLE 2
\begin{deluxetable*}{lccccccccr}
\tabletypesize{\scriptsize}
%\rotate
\tablecaption{\label{complex}Large Sources with Aperture Photometry}
%\tablewidth{0pt}
\tablehead{
\colhead{Source} & \colhead{RA} & \colhead{DEC} & \colhead{F300W} & \colhead{F439W} &
\colhead{F547M} & \colhead{F814W} & \colhead{Age} & \colhead{Log(M$_{Stellar}$)} & \colhead{Area} \\
\colhead{} & degrees & degrees & mag & mag & mag & mag & Myr & {\msun} & kpc$^2$
}
\startdata
Cluster \#1 & 36.27565 & -24.79476 & 18.4$\pm$0.4 & 18.6$\pm$0.4 & 19.8$\pm$0.4 & 19.7$\pm$0.5   & 6$^{+0}_{-2}$ & 5.5$\pm$0.2 & 41.1 \\
Cluster \#2 & 36.27287 & -24.79845 & 17.2$\pm$0.3 & 17.4$\pm$0.3 & 18.8$\pm$0.3 & 18.5$\pm$0.4 & 6$^{+0}_{-2}$ & 6.0$^{+0.1}_{-0.2}$ & 143.9 \\
Nucleus & 36.26851 & -24.78820 & 15.34$\pm$0.04 & 14.88$\pm$0.01 & 16.08$\pm$0.02 & 15.56$\pm$0.03 & 90$^{+0}_{-20}$ & 8.1$^{+0}_{-0.2}$ & 328.3 \\
Tidal Plume & \nodata & \nodata & \nodata & 17.0 & 15.06 & 15.45 & 1300$^{+3700}_{-300}$ & 8.9$\pm$0.2 & 30260.3 \\
%Diffuse Emission & \nodata & \nodata & 9.80 & 9.32 & 9.41 & 9.89 & 250$\pm$40 & 10.8$\pm$0.1 & \nodata \\
\enddata 
\tablecomments{The RA and DEC positions are in J2000 and approximations and are not necessarily related to the exact photometric center of the sources. \\}
\end{deluxetable*}

In an effort to estimate the extinction for each cluster, we used the Q-method applied to cluster colors as described by \citet{john53} and \citet{van68}. This method allows, in principal, an accurate age estimate of the clusters since the Q-Parameters are independent of reddening \citep[see a good example of this method applied to star clusters by][]{whit99}. However, this method requires very good photometry because the error bars get larger when color information is involved. For this reason, and despite the high quality of the WFPC2 images, we did not obtain extinction estimates that are precise enough to apply to each cluster. The ages obtained with the Q-Parameter method were overestimated by $\sim$100\,Myr compared to the more precise SED fitting described previously, which biased the extinction calculation. Note that the overall E(B-V) values we obtained were consistent with the dust features seen by eye in the WFPC2 images, with more obscured zones where larger E(B-V) values were found, but the cluster-to-cluster extinction variation was too large to reliably apply it to each cluster. Although most clusters found in this work are already very young and cannot be that much younger, we must keep in mind that we probably overestimated the age of some clusters, especially the reddest ones. 

\section{The Stellar Cluster Population}
\label{pop}

\subsection{Cluster Luminosity Function}

Figure~\ref{clf} shows the cluster luminosity functions (CLF). For each filter (one per panel) we compare the CLF of four groups. The first group contains all the sources detected in each individual filter and is represented by the black line. The second group, in blue, contains only the clusters from the UBVI sample. We can see that many faint sources are lost when applying the four color detection criteria. The third group, in red, contains the clusters from the BVI sample, i.e., those undetected in the U-band. Several reasons can explain the non-detection of a cluster in the U-band: a) the cluster is old, b) the cluster is faint or c) the cluster is reddened by dust. The CLF in Figure~\ref{clf} clearly shows that most of the sources undetected in the U-band are simply faint, and therefore, probably simply less massive. This is consistent with the young mean and median ages found for this sample as noted in section~\ref{syn}. This is also confirmed by the fact that 59\% of the clusters from the BVI sample have masses smaller than 7$\times$10$^3$\,\msun. The fourth curve in cyan is the CLF for the UBVI and BVI samples combined together. We can see that this combined sample is in better agreement with the raw sample (black) at fainter luminosity compared to the UBVI sample alone (blue).

%% FIGURE 3 %%
\begin{figure*}
\epsscale{0.8}
%\plotone{clftot_cmyk.eps}
\plotone{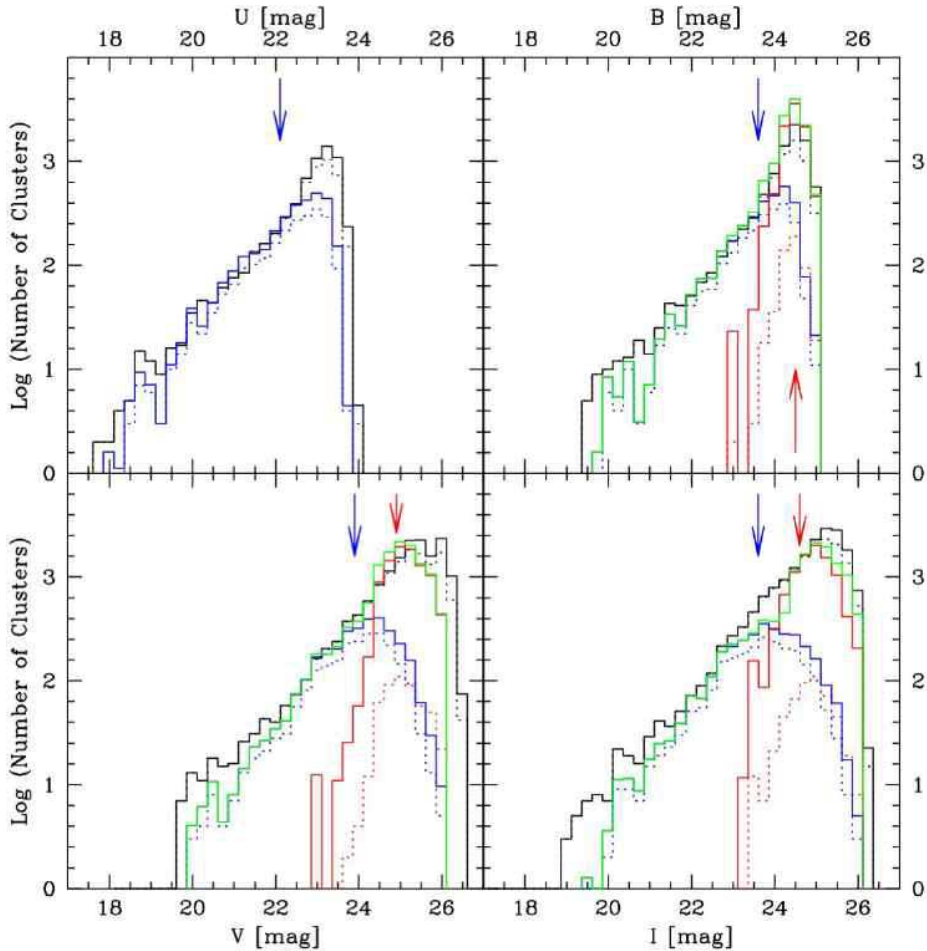}
\caption{\label{clf} Cluster luminosity function for NGC\,922. Black: all sources detected in each individual filter. Blue: Sources detected simultaneously in the UBVI filters. Red: Sources detected simultaneously in the BVI filters only. Green: combination of the UBVI plus BVI samples (i.e. Blue$+$Red curves). The dotted lines represent the samples uncorrected for incompleteness. The arrows show the 50\% completeness limit.}
\end{figure*}

We calculated the slope $\alpha$ of the CLF as described by \citet{elm97}. For the raw samples, we obtained slope values of 2.15$\pm$0.08, 2.22$\pm$0.05, 2.05$\pm$0.05 and 2.10$\pm$0.05 for the U, B, V, and I bands respectively. The slopes tend to steepen to $\sim$2.3 for clusters with UBVI simultaneous detection, when we fit up to the 50\% completeness limit magnitude. Our values are consistent with those of young clusters found in nearby galaxies \citep[e.g.][]{whit99,lar02}.

\subsection{Age Distributions}
\label{distrib}

In Figure~\ref{mapage}, we present spatial maps of clusters for various age ranges. The age group 0-7\,Myr corresponds to clusters capable of ionizing their surrounding gas and therefore producing H$\alpha$ emission. They represent 69\% of the detected clusters, in number, and they are mostly located within the ring of star formation and along the bar, as well as in a more isolated region, South-West of the nucleus. Their location is consistent with the H$\alpha$ emission (Fig.~\ref{cmyk}). We also find that 19\% of the clusters have ages between 7 and 50\,Myr, and only 3\% are older than 100\,Myr.
The oldest clusters are mainly located in the nuclear region, but are also seen along the bar and in the disk. One must keep in mind that some of them might be reddened and therefore younger than estimated here. 

%% FIGURE 4 %%
\begin{figure*}
\epsscale{0.9}
%\plotone{mapage_fourBW.eps}
\plotone{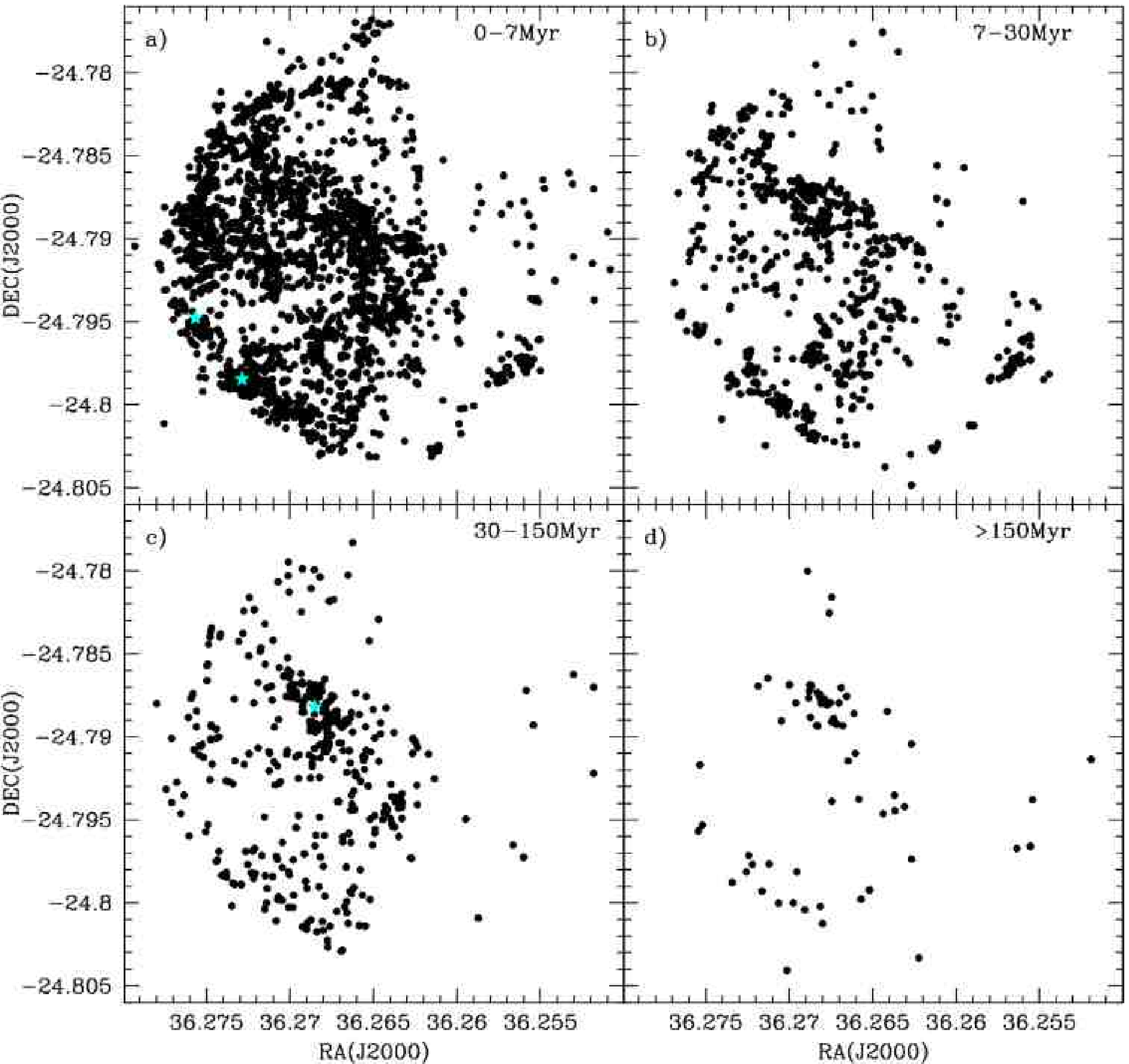}
\caption{\label{mapage} Maps of star clusters detected in NGC\,922 for both UBVI and BVI samples. a) Clusters younger than 7\,Myr and capable of producing H$\alpha$ emission. b) Young clusters with ages from 7 to 30\,Myr. c) Clusters with ages from 30 to 150\,Myr. d) Clusters older than 150\,Myr. The grey star symbols show the location of the two unresolved clusters and the nucleus (see Table~\ref{complex}).}
\end{figure*}

Figure~\ref{lumage} shows the magnitude-age distributions of star clusters in NGC\,922. It can be seen that at older ages ($>$100\,Myr), only the most massive clusters ($>$1$\times$10$^5$\,\msun) can be detected in this galaxy. They are not numerous and they mostly have just BVI photometry. Those sources must be analyzed with care since they are mostly related to the nuclear region, or other very dense regions where the field is very crowded and lumpy, and therefore may include underlying stellar populations. Also, some of them may be particularly affected by extinction and may have ages biased toward larger values, which would lead to overestimated masses. 

%% FIGURE 5 %%
\begin{figure*}
\epsscale{1.0}
%\plotone{lumage.ps}
\plotone{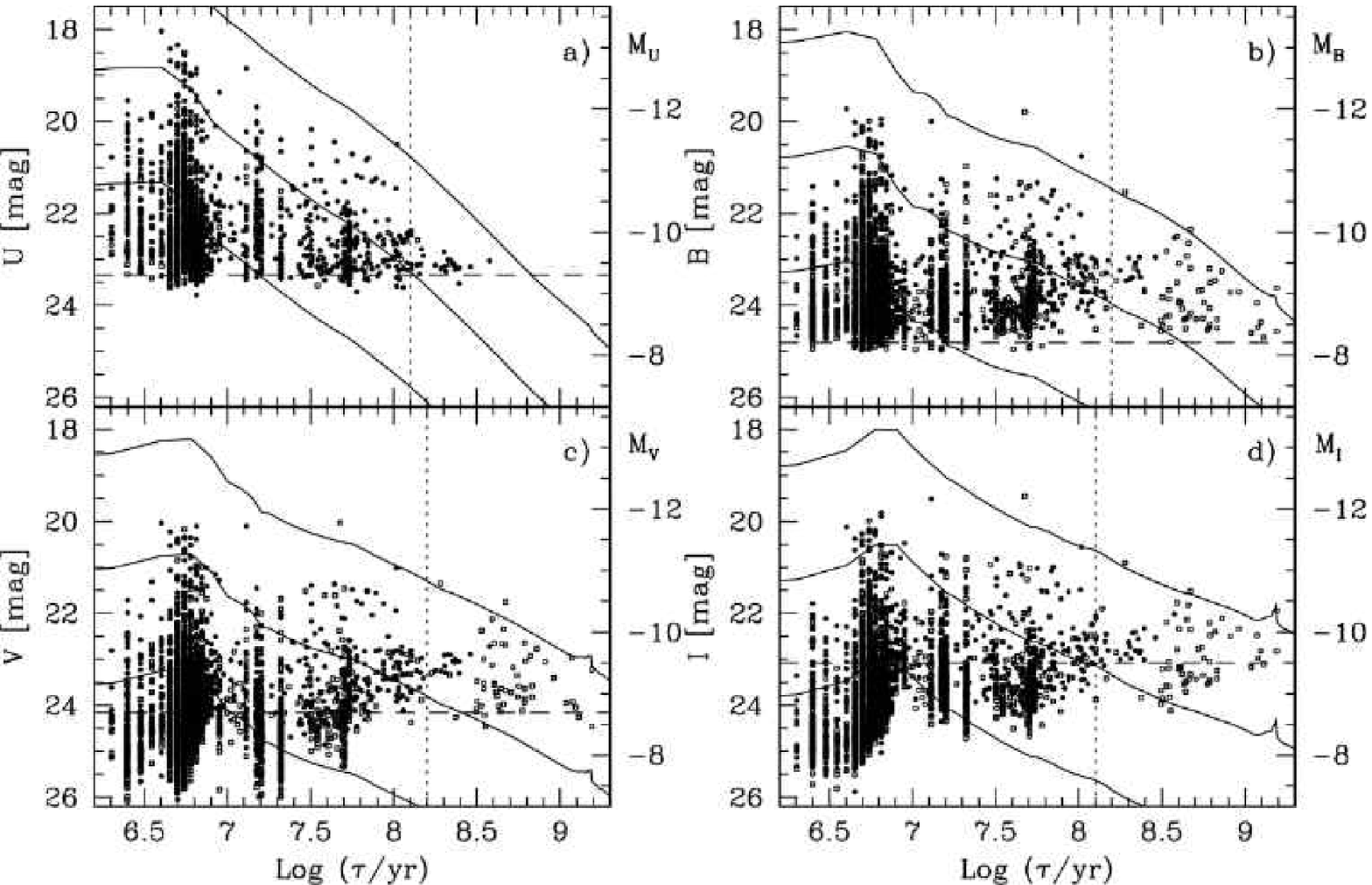}
\caption{\label{lumage} Magnitude-age distributions of star clusters observed in NGC\,922, including both UBVI and BVI samples shown in term of their a) U, b) B, c) V, and d) I magnitudes. Solid circles: clusters with combined completeness level greater or equal to 50\%. Open squares: clusters with combined completeness level smaller than 50\%. The solid lines represent stellar populations with masses of 1$\times$10$^4$, 1$\times$10$^5$, and 1$\times$10$^6$\,\msun, from bottom to top. The vertical dotted lines indicate the age at which the sample of clusters more massive than 1$\times$10$^5$\,\msun\ is complete in the specified filter. The horizontal dashed lines determine the brightness upper limit for a single star in each band. The right axis is displayed in absolute magnitude.}
\end{figure*}

W06 estimate the impact epoch to be $\sim$300\,Myr ago. Because of the small number of clusters at similar ages and their large age uncertainties, we do not believe it is possible to constrain the impact epoch from the cluster age distribution with the current dataset.
Around 10-30\,Myr, there is an effect from the photometric age-dating technique that results in an underestimation of the number of clusters in this age interval. It is commonly claimed as being an artifact from the photometric age-dating technique \citep[e.g.][]{gieles05,lee05} , but more recently \citet{maiz09} proposed that it is more specifically related to stochastic effects in the stellar initial mass function sampling.
In Figure~\ref{lumage} we also indicate for each band the magnitude of the brightest stars, and where confusion can occur between a bright star and a faint cluster. The limits are based on the work of \citet[][their Fig. 4 to 6]{mas06} that includes a large sample of stars in M\,31 and M\,33. In the V band, this is consistent with the work of \citet{hum83}. \citet{hill93} obtained a good stellar photometry in the B-band for massive stars in 30\,Dor and reported a limit value that is slightly fainter than \citet{mas06}. Therefore an averaged limit value was used in this work for the B band. Note that clusters more massive than 1$\times$10$^5$\,\msun\ do not fall below the star/cluster brightness limit for UBV bands.

Age uncertainties are displayed in Figure~\ref{errage} as a function of cluster brightness and age. It shows no particular trend other than larger uncertainties are observed for fainter sources. Square symbols indicates sources for which the completeness level is lower than 50\%. Note that the BVI sample includes only such sources. The bottom panel clearly shows the population of old clusters ($>$300\,Myr) has a low completeness level, and is therefore not reliable to draw scientific conclusions. 

%% FIGURE 6 %%
\begin{figure*}
\epsscale{0.7}
%\plotone{errage.ps}
\plotone{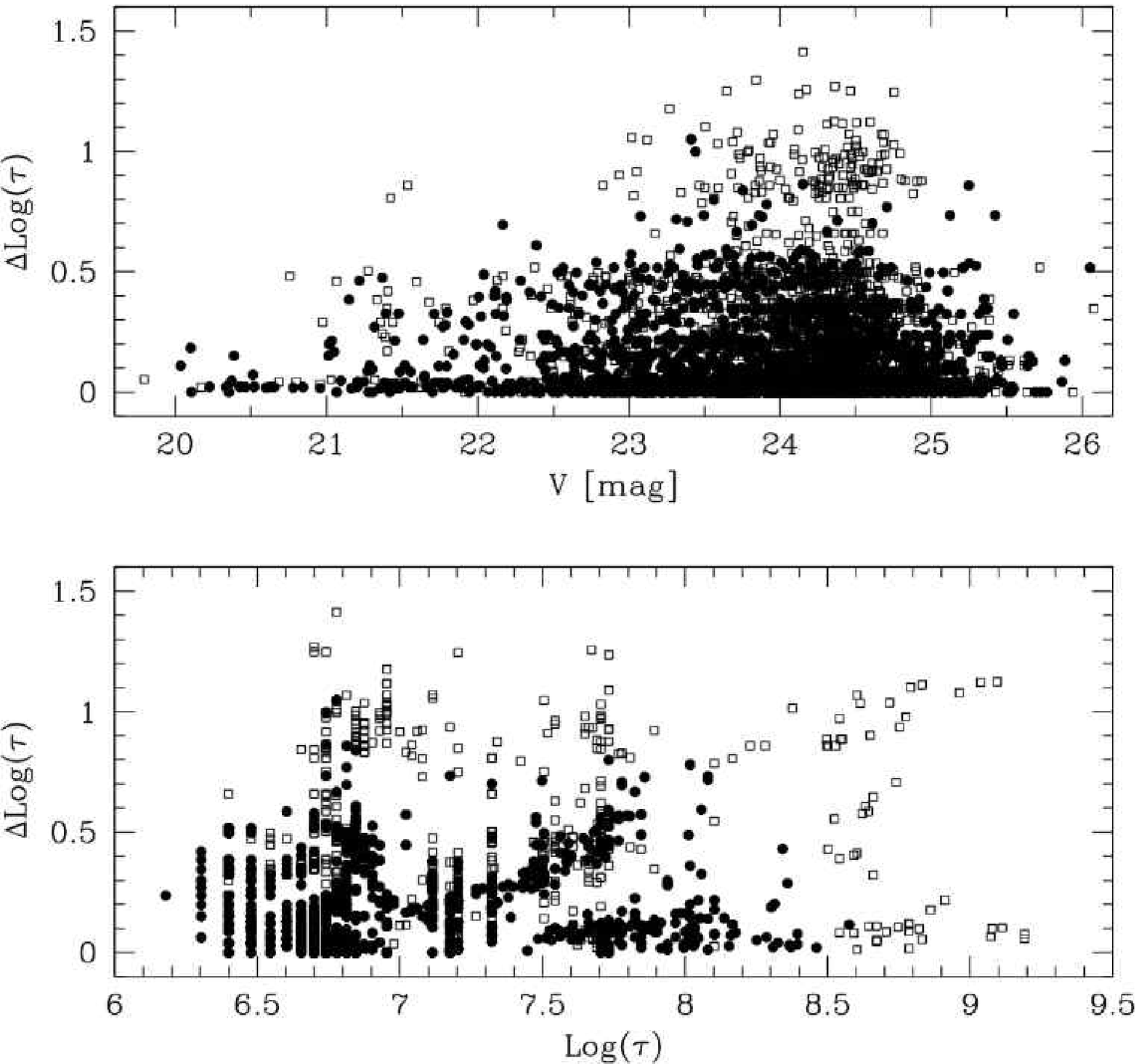}
\caption{\label{errage} Top panel: Age uncertainties as a function of cluster brightness. Bottom panel: age uncertainties as a function of cluster age. Solid circles: clusters with completeness level greater or equal to 50\%. Open squares: clusters with completeness level smaller than 50\%. }
\end{figure*}

To determine the age at which the sample is complete, we use Figure~\ref{lumage} and find the age where the model lines (solid) cross the lower envelope of the data points with $>$50\%\ completeness and the horizontal (dashed) line marking the brightest stars. The lowest age of these two determines the age completeness for that filter and mass limit, and is marked by a dotted line in each panel of  Figure~\ref{lumage}; the lowest of these for U, B, V, and I determines our final age limit, since clusters must be detected in at least these filters to have a photometric age. This results in an age limit of 125\,Myr for the 1$\times$10$^5$\,\msun\ sample which is effectively defined by both the brightest star luminosity and the 1$\times$10$^5$\,\msun\ model line in the U band.  
Those values indicate that we do not have to worry about statistical fluctuations of the initial mass function that can be seen for clusters less massive than a few times 10$^4$\,\msun\, \citep{cer04}. Also, it means that we cannot extend our mass-limited sample analysis to ages greater than $\sim$125\,Myr. 

The age distribution of stellar clusters in NGC\,922 is presented in Figure~\ref{agedist}. A detailed list of the completeness corrections applied to Fig.~\ref{agedist} is presented in Table~\ref{tabcomp}. The age interval from 1 to 330\,Myr covers essentially the time over which NGC\,922 should have experienced a relatively high star formation activity and produced a large amount of star clusters due to the collision with S2, and the subsequent propagation of the density wave through the disk (W06). The number of clusters was corrected for incompleteness (see \S\ref{data}), and the age distributions include only sources with combined completeness levels higher than 50\%. 

%% FIGURE 7 %%
\begin{figure*}[!t]
\epsscale{0.9}
%\plotone{agedistot.ps}
\plotone{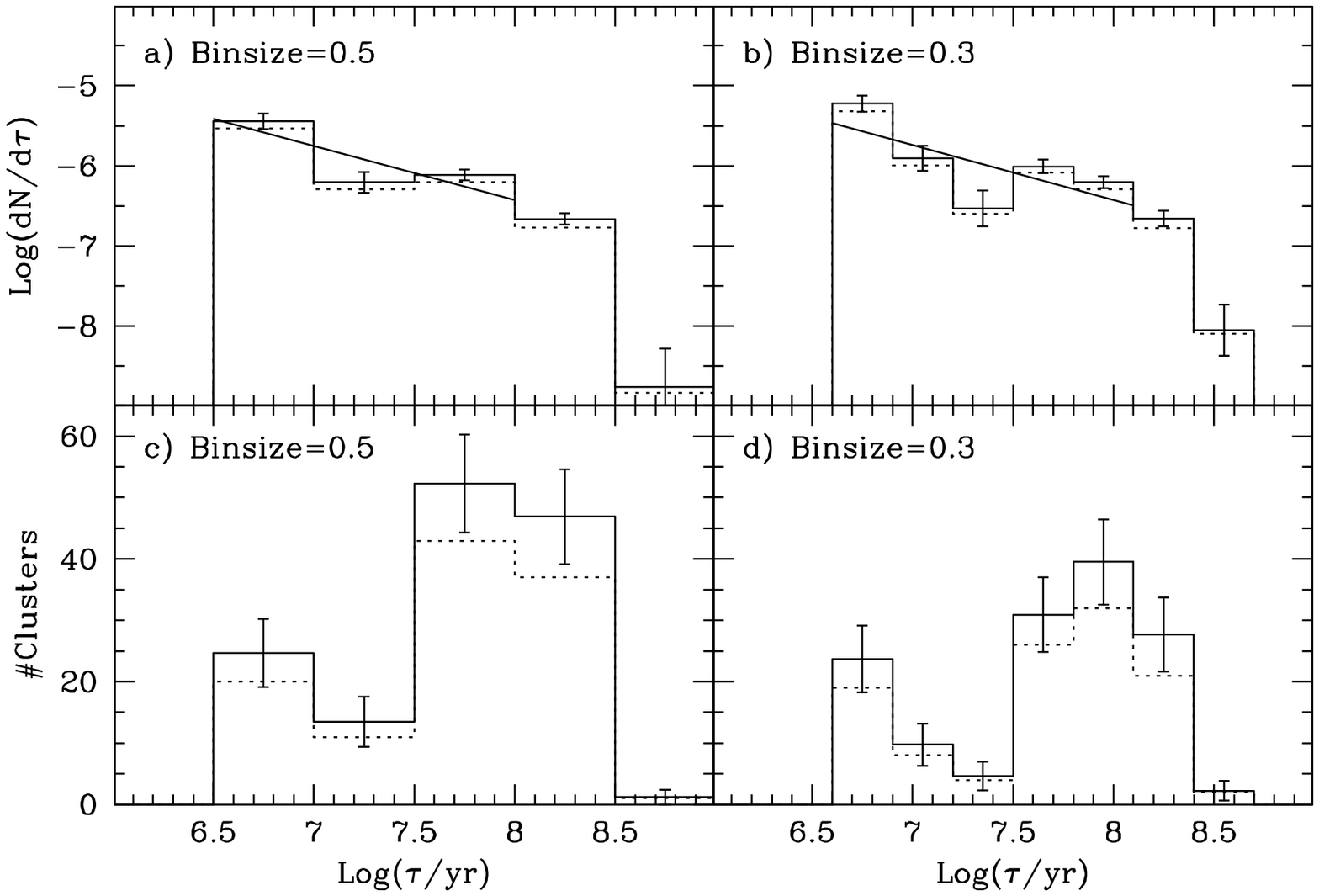}
\caption{\label{agedist} a) Observed age distribution of star clusters of at least 1$\times$10$^5$\,\msun\ using a bin size of 0.5 dex in log(age). b) Same as in panel a) but for a bin size of 0.3 dex. Linear fits for 6.5$\ge$Log($\tau$/yr)$\ge$8.0, where the sample is considered as complete (see \S\ref{distrib}), are over-plotted. c) Simple cluster age histograms for the same sample using a bin size of 0.5 dex. d) Same as in panel c) but using a bin size of 0.3 dex. Dotted lines: sample uncorrected for incompleteness. Solid lines: sample corrected for incompleteness. Clusters with completeness lower than 50\%\ are excluded.\\}
\end{figure*}

%TABLE 3
\begin{deluxetable*}{lccc}
\tabletypesize{\scriptsize}
\tablecaption{\label{tabcomp} Incompleteness Corrections of Observed Cluster Numbers and Age Distribution Values}
\tablehead{
 \colhead{Age Bin} & \colhead{No. Clusters\tablenotemark{a}} & \colhead{No. Cluster\tablenotemark{b}} & \colhead{Log(dN/dt)} \\
 \colhead{} & \colhead{Uncorrected} & \colhead{Corrected} & \colhead{Corrected}}
\startdata
0.3\,dex bins & & & \\
6.6-6.9 &  19.0$\pm$4.3 &  23.7$\pm$5.4 &  -5.22 \\
6.9-7.2  &  8.0$\pm$2.8  &  9.8$\pm$3.5 &  -5.91 \\
7.2-7.5  &  4.0$\pm$2.0  &  4.7$\pm$2.3 &  -6.53 \\
7.5-7.8 & 26.0$\pm$5.1 &  30.9$\pm$6.1 &  -6.01 \\
7.8-8.1 &  32.0$\pm$5.7 &  39.5$\pm$7.0 &  -6.20 \\
\hline
0.5\,dex bins & & & \\
6.5-7.0 & 20.0$\pm$4.5 &  24.7$\pm$5.5 & -5.44 \\
7.0-7.5  & 11.0$\pm$3.3  &  13.5$\pm$4.1 & -6.21 \\
7.5-8.0  &  43.0$\pm$6.5 & 52.3$\pm$8.0 & -6.12 \\
\enddata 
\tablenotetext{a}{Number of clusters observed and uncorrected for incompleteness. The values corresponds to the dotted lines in Fig.~\ref{agedist}c,d.}
\tablenotetext{b}{Number of clusters observed and corrected for incompleteness. The values corresponds to the solid lines in Fig.~\ref{agedist}c,d.}
\end{deluxetable*}

We find slopes of $-$0.7$\pm$0.2 for the age distributions in Figure~\ref{agedist}. The slopes were obtained using ordinary least-squares of Y$|$X linear regressions \citep{fei92} over the mass ranges that are not affected by incompleteness, i.e. 6.5\,$<log(\tau)<$\,8.0 if using a bin size of 0.5 dex and 6.6\,$<log(\tau)<$\,8.1 if using a bin size of 0.3 dex. Note that a bin size of 0.3 dex is smaller than the typical age uncertainties as shown in Fig~\ref{errage}. We excluded the first bins (0-5\,Myr) because clusters may still be buried in their native dust cocoon at this young age, adding to the incompleteness at visible wavelengths. We also excluded the bin around 30\,Myr in Figure~\ref{agedist}b due to the age bias discussed above. The 30\,Myr bin was included in the fit in Figure~\ref{agedist}a because of the very low number of bins available for the fit. 
We also performed a Kolmogorov-Smirnov (KS) test as an independent test to check the slope of the observed age distribution. The test is presented in Fig.~\ref{kstest} and show results consistent with negative slopes. We find a probability of less than 1\% for a positive slope, while we obtain the highest probabilities ($>$50\%) for slopes between $-$0.4 and $-$0.8. Note that due to jumps in the cumulative distribution, we cannot find a single best fit.

The evolution of the SFR as traced by star clusters should, to first order, tell us something about the distribution of the dense  interstellar medium (ISM), or more specifically the molecular gas, from which the clusters formed.  In a collisional ring galaxy much of the star formation  occurs in the expanding ring.  Hence, one may expect the SFR to be proportional to the rate of the dense ISM being swept up by this ring.  If the expansion velocity is nearly constant, and the ring radius is significantly larger than the radius at which the impact occurred, and if there is no cluster dissolution then a flat cluster age distribution would mean that the dense ISM must have a surface density $\Sigma_g$ that falls off with radius, $r$, as $r^{-1}$.  While the neutral ISM in galaxies can fall off this shallowly \citep{bos81, hoek01,leroy08}, the molecular ISM, from which stars form, has a steeper fall-off more similar to the stellar distribution \citep{leroy08}.  This will produce positive slopes in $\log(N)$ versus $\log({\rm age})$, as we indeed find in the simulations presented in Section~6.  A negative slope suggests a shallower drop-off than $\Sigma_g(r) \propto r^{-1}$; while one can not rule this out in any particular galaxy, it would represent a peculiar molecular ISM distribution.  Assuming a normal steep $\Sigma_g(r)$ profile, one way to reconcile the negative slope observed in NGC\,922 with what we understand of collisional ring galaxies is to introduce young star cluster disruption and dissolution. In the next section we explore this possibility.

\section{Star Cluster Disruption in NGC\,922}
\label{test}

%% FIGURE 8 %%
\begin{figure*}
\epsscale{0.8}
%\plotone{kstestraw_1E5msun.ps}
\plotone{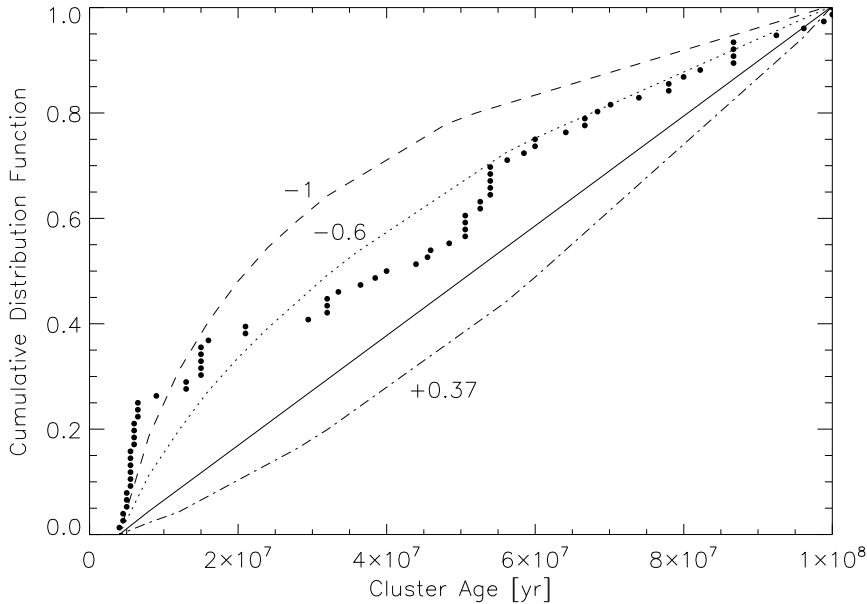}
\caption{\label{kstest} Cumulative distribution function of star clusters of at least 1$\times$10$^5$\,\msun\ (circles). For comparison, the dashed, dotted, plain, and dash-dotted lines show what would be the cumulative distribution function of an age distribution with slopes of $-$1, $-$0.6, 0, and $+$0.37, respectively. The positive slope represents a typical case of a ring galaxy the size of NGC\,922 as simulated in \S\ref{simresult}.\\}
\end{figure*}

The disruption of star clusters is a concept not trivial to test observationally, especially because the star formation history is often unknown. The recent work of \citet{bas09} expresses well this problem. NGC\,922 represents a good case-study for the star cluster disruption scenario since its cluster formation history can be relatively well recovered with dynamical and star formation simulations.

In their work on the Antennae galaxies, \citet{fall05} proposed that about 90\% of star clusters dissolve for each dex of age, up to 1\,Gyr, independently of cluster mass. We will call this empirically based relation the Mass Independent (MI) scenario, following the works of \citet{fall05} and \citet{whit07}. Concerns about the mass independent, and continuous nature of the cluster destruction, as well as the assumption of a constant SFR have been raised \citep[e.g.][]{bas09}. 
There is an alternative scenario in which cluster dissolution rate varies over time. According to this scenario, infant mortality is responsible for cluster dissolution during the first 10$^7$\,yr or so, leading to a steep decreasing slope in the age-distribution. Infant mortality of star clusters is a phenomenon thought to happen to young clusters experiencing a serious loss of gas and dust during the supernovae explosion phase, leaving the clusters gravitationally unbound and easy to disrupt within 10 to 50\,Myr \citep[e.g.][]{lada03,fall05,good08}. 
According to a recent work, infant mortality is hardly seen at visible wavelengths \citep{gieles08}.
Then internal dynamical evolution and stellar population fading occurs between 10$^7$ and 10$^8$\,yr. 
There are relatively few cluster destroyed during that phase. This results in a flat age distribution as sketched by Lamers (2008), although the precise form may not be exactly flat, which he also notes.
Finally mass dependent cluster dissolution, due to galactic tidal effects and other external effects, produces a decreasing slope for the 10$^8$-10$^9$\,yr. We will call this scenario the Mass Dependent (MD) scenario, after the works of \citet{bou03}, \citet{gieles05}, \citet{lamers05}, \citet{lamers06}, and \citet{gieles07}. 
A good comparison between the two scenarios is presented by \citet{lamers08}. As shown in Figure~\ref{agedist}, there might be some evidence of a flat distribution in NGC\,922 between 10 and 100\,Myr, as proposed by the MD scenario, but that could also be due to the artificial lack of clusters around 10-30\,Myr mentioned above. 

For every cluster detected in NGC\,922, we applied a correction for the loss of stellar clusters due to dissolution. 
We used the two cluster disruption functions discussed above.  A starting age of 4\,Myr was chosen instead of 0\,Myr because it corresponds better to the age of the first supernova explosions and, presumably, the time when the clusters can start dissolving. The fraction of star clusters F$_{sc}$ left after dissolution, according to the MI model, is then given by

\begin{equation}
log(\rm{F}_{sc})=1,~for~log(\tau)\leq6.6
\end{equation}
\begin{equation}
log(\rm{F}_{sc})=6.6-log(\tau),~for~log(\tau)>6.6
\end{equation}

\noindent
where $\tau$ is the age of the cluster in years. Following the MD model, the fraction can be expressed by 

\begin{equation}
log(\rm{F}_{sc})=1,~for~log(\tau)\leq6.6
\end{equation}
\begin{equation}
log(\rm{F}_{sc})=16.5-2.5log(\tau),~for~6.6<log(\tau)<7.0
\end{equation}
\begin{equation}
log(\rm{F}_{sc})=-1,~for~7.0<log(\tau)<8.0
\end{equation}
\begin{equation}
log(\rm{F}_{sc})=12.6-1.7log(\tau),~for~log(\tau)>8.0
\end{equation}

For the early ages of the MD model, we used a 90\% infant mortality rate per decade of time.
Note that because we do not know the star formation history of NGC\,922 prior to the collision, we cannot discuss the issue of whether or not the cluster disappearance in this age range is mass-dependent (see \S\ref{distrib}).

The age distributions of star clusters corrected for cluster dissolution and incompleteness is presented in Figure~\ref{agedistback}. We find positive slopes of $+$0.5$\pm$0.2 for clusters with masses of at least 1$\times$10$^5$\,\msun\ when we use the MI model. We find a relatively flat slopes (0.0$\pm$0.1) when we use the MD model. The slopes were calculated up to $log(\tau)=$ 8.1. 

%% FIGURE 9 %%
\begin{figure*}
\epsscale{0.9}
%\plotone{agedistback6.ps}
\plotone{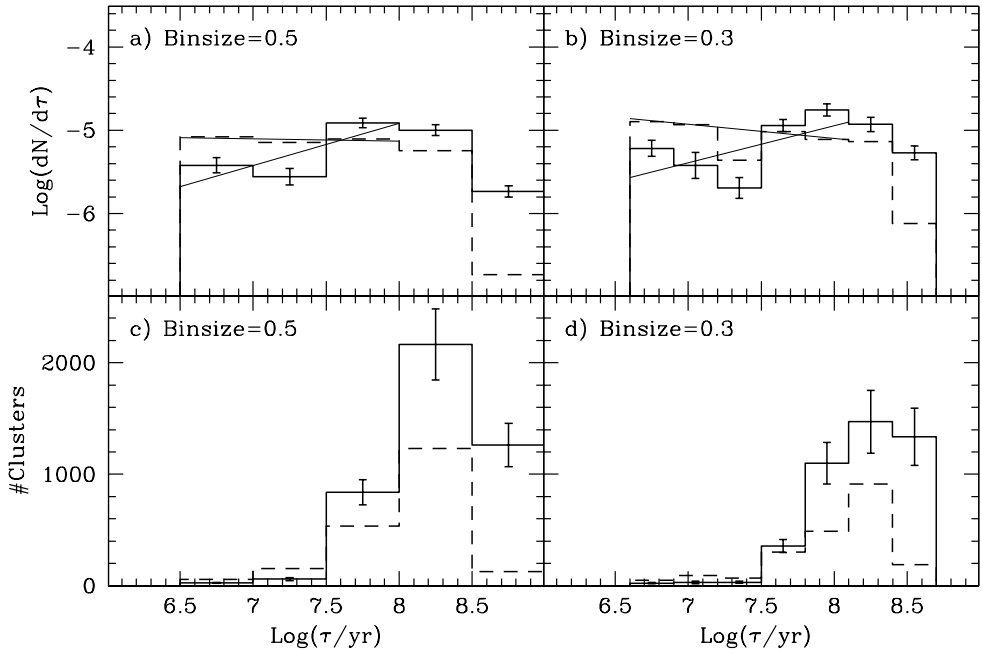}
\caption{\label{agedistback} a)  Observed age distribution of star clusters of at least 1$\times$10$^5$\,\msun\ and corrected for cluster disruption and incompleteness. b) Same as in panel a) but for a 0.3\,dex bin size. c) Simple cluster age histogram for the sample corrected for cluster disruption and incompleteness. d) Same as in panel c) but for a different bin size. The solid lines show the sample corrected using the MI model while the dashed lines show the sample corrected using the MD model. The error bars on the MD corrected histograms are omitted here so that the plots are not too cluttered. They are similar in size to those for the MI corrected histograms.\\ }
\end{figure*}

The positive slope value below 100\,Myr now suggests that the cluster formation rate was decreasing in NGC\,922 during that period, assuming that the dissolution correction is realistic and that there are no other major mechanisms affecting the number of clusters. The slope value for clusters older than 100\,Myr is still steep despite the correction for dissolution but this is attributed to incompleteness. 

\section{Dynamical and Star-Cluster Simulations}
\label{model}

In order to better understand the age distribution of star clusters observed in NGC\,922, we performed N-body/SPH simulations on possible formation sites of star clusters in NGC\,922 colliding with its companion S2. We considered that the observed ring-like morphology of NGC\,922 is the result of a high-speed, off-centered collision between NGC\,922 and S2 (W06). We therefore describe the simulation results with unique orbital configurations and disk inclinations. Here we focus mainly on the age and spatial distributions of star clusters within NGC\,922 and not the chemical and morphological evolution. 

\subsection{Model Parameters}

In the current simulation NGC\,922 is represented by a fully self-consistent disk galaxy model whereas S2 is represented by a point mass. NGC\,922 is modeled as a disk galaxy embedded in a massive dark matter halo with a total mass of $M_{\rm dm}$ and the universal ``NFW profile'' \citep{nav96}. The disk has an exponential radius and the total disk mass and size are $M_{\rm d}$ and $R_{\rm d}$, respectively. Our results were obtained using $M_{\rm d}=$2$\times$10$^{10}$\,\msun, $R_{\rm d}=13.4$\,kpc, and $M_{\rm dm}/M_{\rm d}=9$. The gas mass fraction $f_{\rm g}$ of the disk is assumed to be a free parameter.
More details about the numerical methods and techniques we employ for modeling chemodynamical evolution of interacting galaxies with star formation can by found in \citet{bekki02}.  

The radial ($R$) and vertical ($Z$) density profiles of the disk are assumed to be proportional to $\exp (-R/R_{0}) $ with scale length $R_{0} = 0.2R_{\rm d}$ and to ${\rm sech}^2 (Z/Z_{0})$ with scale length $Z_{0} = 0.04R_{\rm d}$, respectively. In addition to the rotational velocity attributable to the gravitational field of the disk and halo components, the initial radial and azimuthal velocity dispersions are added to the disk component in accordance with the epicyclic theory, and with a Toomre Q parameter value of 1.5 \citep{bin87}. The interstellar medium in NGC 922 is represented by SPH particles. The mass resolution (i.e., the minimum mass of a particle) and the size resolution (the minimum gravitational softening length) are 2$\times$10$^5$\,\msun\ and $13.4$\,pc, respectively.

The mass ratio of the companion dwarf to the spiral ($m_{\rm 2}$) is set to be 0.2. The initial location and velocity of the companion S2, with respect to be the center of the spiral, are represented by ${\bf X}_{\rm g}$ and ${\bf V}_{\rm g}$, respectively. The spiral is assumed to be inclined by 80 degrees with respect to the $x$-$y$ plane. Although we have investigated models with different ${\bf X}_{\rm g}$ and ${\bf V}_{\rm g}$, we show the results of the model with ($x$,$y$,$z$) = ($-4R_{\rm d}$, 0.2$R_{\rm d}$, 0) and ($v_{\rm x}$,$v_{\rm y}$,$v_{\rm z}$) = ($6v$, 0, 0), where $v$ is the simulation unit ($=80.1$ km s$^{-1}$). Stars that are initially within the spiral disk are referred to  as ``stars'' (or ``old stars'') whereas stars that are newly formed from gas are  as ``new stars''. 

\subsection{Possible formation sites of star clusters}

We adopted the following star formation laws to convert gaseous particles into either star clusters  or field stars. For star cluster formation, we considered that interstellar gaseous pressure ($P_{\rm gas}$) in star forming regions of a galaxy drives the collapse of pressure-confined magnetized self-gravitating molecular clouds to form compact clusters, providing that $P_{\rm gas}$ is larger than the surface pressure ($P_{\rm s}$) of the clouds \citep[e.g.][]{elm97}:

\begin{equation}
P_{\rm gas} \ge  P_{\rm s} \sim 2.0\times 10^5 k_{\rm B}. \;
\end{equation}  

\noindent
where $k_B$ is the Boltzman constant. This adopted value for the threshold gaseous pressure $P_{\rm gas}$ would be appropriate for the formation of massive star clusters, such as the 1$\times$10$^5$\,\msun\ mass-limited sample. Our star formation prescription allows for the formation of low mass clusters, associations or field stars even if $P_{\rm gas}$ is smaller than the above threshold value. However, we do not track these stars in the plots presented here.

For field stars, we adopt the Schmidt law \citep{sch59} with an exponent of 1.5 \citep{ken89}. Although we cannot investigate the detailed physical processes of cluster formation in the present and global simulation (i.e. from $\sim$ 100 pc to 10 kpc scale), we expect that the adopted phenomenological approach enables us to identify the plausible formation sites of star clusters in NGC 922. For models with $f_{\rm g}=0.4$ (note that this is different from the 0.2 value adopted by W06), the present possible mass of young stars in NGC\,922 can be well reproduced. For the adopted orbital configurations, NGC\,922 can experience a moderately strong starburst when S2 passes through NGC\,922.

\subsection{Simulation Results}
\label{simresult}

The best-fit model produces a peak of star formation during the passage of S2 within the disk of NGC\,922 about 150\,Myr ago, then a relatively constant decrease of the cluster formation rate over time. Figure~\ref{simhist} shows the simulated age distribution of 4240 star clusters formed as described above \citep[see a detailed discussion about the star cluster or globular cluster formation models by][]{bekki02}. The overall shape of the simulated age distribution displays a slope of $+$0.37$\pm$0.02 between 1 and 100\,Myr. 

%% FIGURE 10 %%
\begin{figure*}
\epsscale{0.8}
%\plotone{simf9.ps}
\plotone{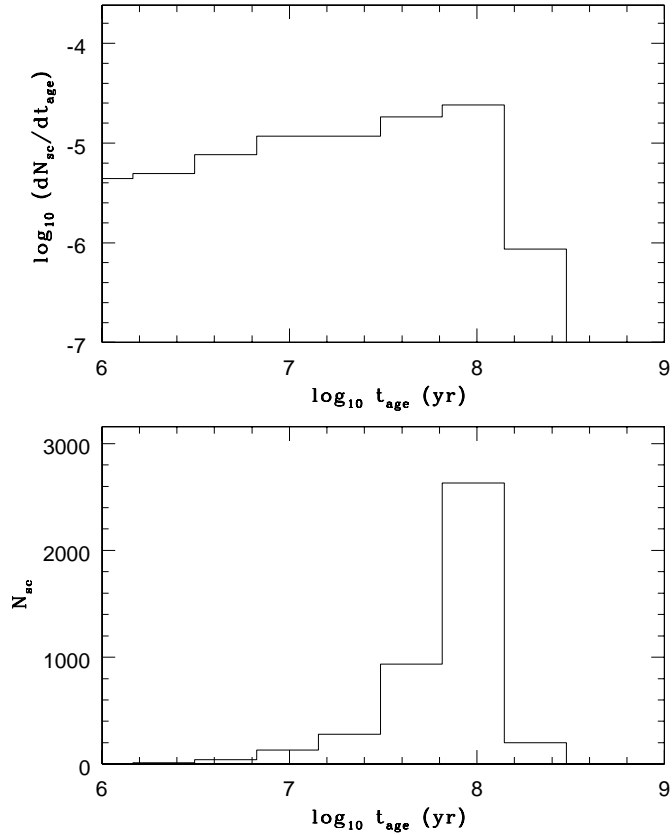}
\caption{\label{simhist} Top: simulated age distribution of clusters formed after NGC\,922's collision with S2. Bottom: Simulated cluster age histogram.}
\end{figure*}

The expected cumulative distribution function of clusters formed with this age distribution and no disruption is shown in Fig.~\ref{kstest} with a dot-dashed line. It shows that the age slope of $-0.7 \pm 0.2$ we find without any cluster dissolution correction is significantly steeper than our simulation result by $\sim5\sigma$.  As discussed in Section~\ref{test} and illustrated in Fig.~\ref{agedistback}, after correction for MI we find an age slope of $0.5 \pm 0.2$ which is within $\sim 1\sigma$ of the simulated results. Correction using the MD scenario produces a flatter age slope of $0.0 \pm 0.1$, which differs from our simulation by $\sim4\sigma$.  We therefore conclude that correction for cluster dissolution, using previously defined prescriptions, is capable of bringing the observed age distribution into better accord with the simulations, and that the MI model for the correction produces results somewhat more consistent with the simulations than the MD based correction. Because of the possibility that any given galaxy may have a peculiar cluster formation history that is not captured in a simple model, one should not see our results as a sweeping proof of cluster destruction following the MI scenario, the MD scenario still agrees within 4$\sigma$ of the simple model.  However, it does suggest that further comparisons between simulations and well modelled galaxies may provide good constraints on which cluster disruption path works best.

One possible problem with the MI scenario is the vast number of clusters that would have had to be destroyed. To test if the implied SFR is consistent with the observations of W06, we integrated the MI cluster formation rate for all detected clusters uncorrected for incompleteness.
If we use the MI scenario, we obtain a total stellar mass post-collision of 7$\times$10$^8$\,\msun. This is 14\%\ of the total stellar mass reported by W06 (5$\times$10$^9$\,\msun) for the whole galaxy. This stellar mass is a lower limit if we consider that the collision was 300\,Myr ago as suggested by W06, and that our sample is not complete beyond 125\,Myr ago. However, we found an
error in the conversion of the 2MASS fluxes to ABmag presented by W06. We re-calculated the photometry and masses (see Table~\ref{newphot}) and obtain a total stellar mass of 2.8$\times$10$^{10}$\,\msun\ for NGC\,922 based on the K-band, which makes the corrected mass originally created in clusters to be 2.5\%\ of the total stellar mass. We concluded that the integrated cluster mass produced in this starburst is not an unreasonable fraction of the total mass.

%TABLE 4
\begin{deluxetable}{lcc}[!h]
\tabletypesize{\scriptsize}
\tablecaption{\label{newphot} Corrected 2MASS Photometry for NGC\,922 and S2}
\tablehead{
 \colhead{} & \colhead{NGC\,922} & \colhead{S2}}
\startdata
NUV (AB mag) & 13.04 & 15.98 \\
R (AB mag) & 11.57 & 14.71 \\
J (ABmag) & 11.57 & 14.87 \\
H (AB mag) & 11.49 & 14.89 \\
K (AB mag) & 11.82 & 14.80 \\
M$_{\star}$ (\msun) & 2.8$\times$10$^{10}$ & 1.81$\times$10$^9$ \\
\enddata 
\tablecomments{All values have been corrected for extinction as in W06.}
\end{deluxetable}

%%FIGURE 11 %%
\begin{figure*}[!t]
\epsscale{0.72} 
%\plotone{clumpsID.ps}
\plotone{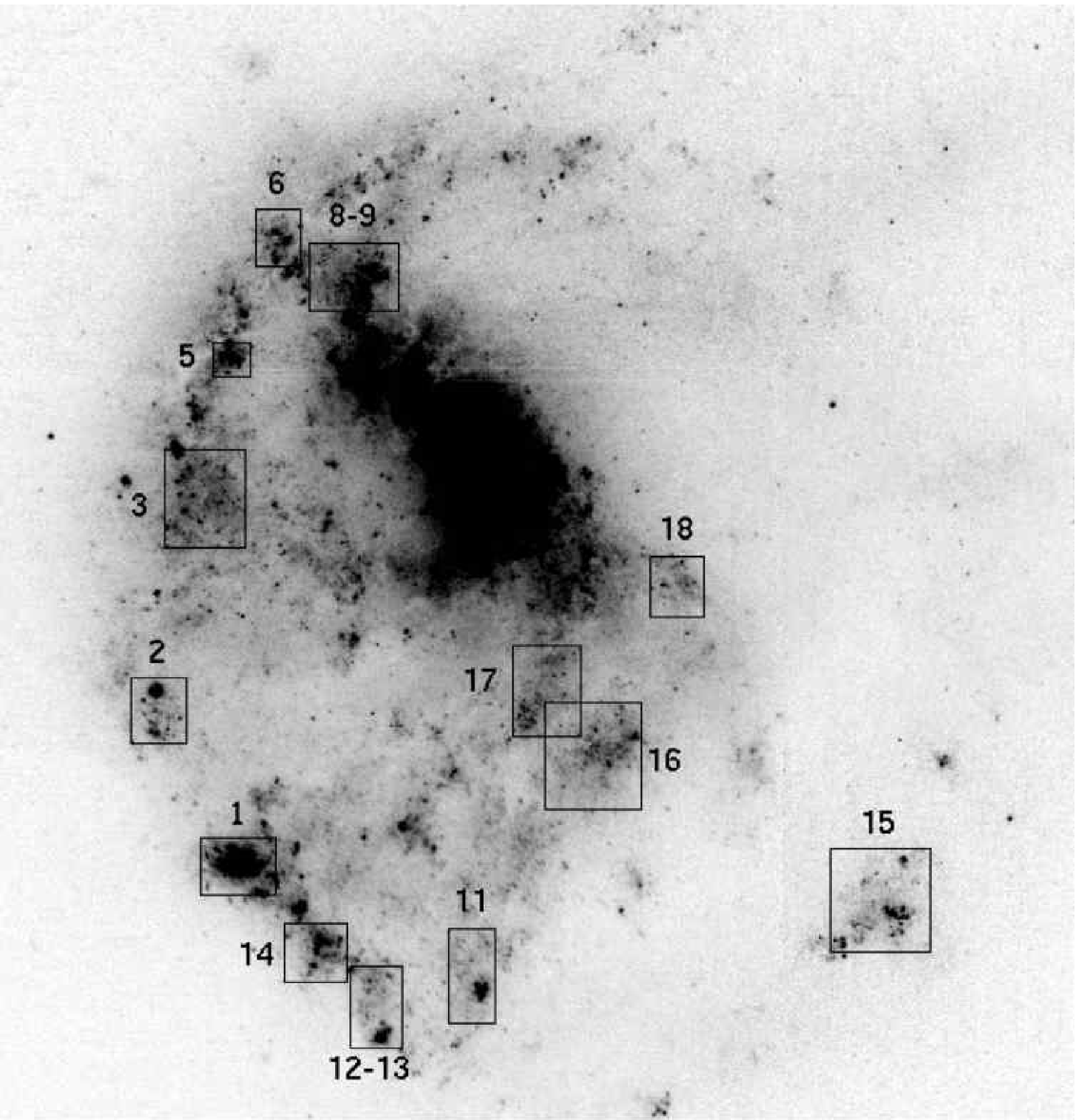}
\caption{Large-scale star-forming complexes are labeled on the F547M image of NGC 922.}\label{figbig}
\end{figure*}

The simulated star clusters younger than 100\,Myr are located preferentially in the ring. This result confirms that the high-density, high-pressure ring formed after the NGC\,922-S2 collision is the ideal site for star cluster formation, and is consistent with the observations.
Also, according to the present simulation, NGC\,922 and the companion collided about 150\,Myr ago which triggered a starburst in the central region of NGC\,922. This burst population does not show a ring-like distribution and is thus responsible for a more dispersed distribution for star clusters with ages between 100 and 200 Myr. This population is not obvious in the HST images, but this is related to the detection limit of our sample or cluster dissolution, most of the older clusters being seen either within the nuclear region or the ring. However, in contradiction to the models, the observations show that star clusters also formed inside the ring within the last 100\,Myr. Those were possibly formed along some spoke features, as shown in other collisional ring simulations \citep{hern93,apple96,struck97}.

\section{Large-Scale Star Formation}
\label{large}

Large-scale star formation in galaxies provides insight into global dynamical processes. In normal spiral galaxies, the largest star-forming units are kpc-size complexes which incorporate many smaller clumps or clusters.  The sizes of these largest complexes scale with the absolute magnitude and the diameter of the galaxy \citep[][hereafter EELS]{eels}. They are governed by a Jeans instability \citep{elm94a,elm94b}, which often leads to rather regular spacing along spiral arms \citep{elm83} and in circumnuclear rings \citep{buta93,maoz96,elmetal97}. The gas mass of H\,{\sc{i}} supercloud complexes is on the order of $10^7$\,\msun\ for spiral arms \citep{elm87} and the total stellar mass upper limit is less than $10^6$\,\msun\, consisting of several embedded clusters; in circumnuclear rings, the hotspot stellar masses are typically a few times $10^5$\,\msun\ \citep{elmetal97}. Complexes are also the smallest structures that can be resolved in high redshift galaxies, so these star-forming units are useful to examine in NGC 922 for comparison with ring galaxies in the early universe.

\subsection{Star-forming complexes}
\label{complexes} 

Complexes in NGC 922 were identified by visual inspection of the WFPC2 images. Figure \ref{figbig} shows the F547M image with the star-forming complexes labeled; these encompass the smaller-scale clusters discussed in sections~\ref{pop} to \ref{model}. These complexes, which range from 750 pc to 2 kpc in diameter, are also prominent in the SINGG H$\alpha$ image of NGC 922 (see Fig.~\ref{cmyk}). In nearby quiescent spiral galaxies, a late-type galaxy with NGC 922's absolute magnitude of M$_B$=-21.24 \citep{meurer06} would be expected to have a maximum complex size of $\sim$1\,kpc (see EELS).

Photometry was done on the complexes in each of the WFPC2 images by defining boxes around the complexes and their brightest subunits and using the IRAF task {\it imstat}. The results were compared with evolutionary models of an instantaneous burst, as in section 3. The stellar population synthesis results are presented in Table~\ref{synclump}. The complexes have ages ranging from 4 to 86\,Myr and masses from 9$\times$10$^5$ to 3$\times$10$^7$\,\msun. 
%Figure \ref{figcol} shows a plot of the (B-V) vs. (V-I)  colors for the clumps. Black dots indicate the brightest subunits of the complexes, while blue dots represent the complexes (Figure \ref{figbig}). %The line is an evolutionary model from Starburst99 (Leitherer et al. 1999), taken from the online dataset.  NGC 922 has solar metallicity (Wong et al. 2006), so we used the z=0.02 model. Because these are star-forming regions, we used the instantaneous curve, and selected the Salpeter mass function with an upper mass limit of 30 M$_{\odot}$. Log (age) is indicated along the evolutionary track by numbers. The small subunits have slightly younger ages than the whole complexes, as expected if star formation takes place over a period of time throughout the complex. The subunits have log (age) of about 7.3 to 8.2, while the complexes have log (age) of about 7.8 to 8.6. 

%Using these approximate ages, we can estimate masses from a color-magnitude diagram. This is shown in Figure \ref{figmag} with absolute V magnitude versus (V-I), using the distance of 43 Mpc (Wong et al. 2006) to determine the absolute magnitudes of the complexes. This figure just shows the complex subunits, along with three evolutionary tracks for the model used in the color-color figure but for 3 different masses. The black line is for a cluster of $10^6$ M$_{\odot}$, while the green is for $10^7$ M$_{\odot}$. The red line is the best fit for the subunits for log (age) around 7.5, representing a mass of 3.6$\times$ $10^6$ M$_{\odot}$. (The brightest clump is younger than 10$^7$ years.)

%TABLE 5
\begin{deluxetable*}{cccccccccc}
\tabletypesize{\scriptsize}
\tablecaption{\label{synclump} Physical Parameters of Star-Forming complexes}
\tablehead{
\colhead{Complex ID} & \colhead{RA} & \colhead{DEC} & \colhead{F300W} & \colhead{F439W} &
\colhead{F547M} & \colhead{F814W} & \colhead{Age} & \colhead{Log(M$_{Stellar}$)} & \colhead{Diameter} \\
\colhead{} & degrees & degrees & mag & mag & mag & mag & Myr & {\msun} & kpc
}
\startdata
1 & 36.272875 & -24.798485 & 15.49$\pm$0.07 & 16.05$\pm$0.07 & 16.6$\pm$0.1 & 17.4$\pm$0.2 & 4$^{+50}_{-4}$ & 6.7$^{+1.1}_{-0.3}$ & 0.76 \\
2 & 36.275468 & -24.795138 & 16.72$\pm$0.04 & 17.01$\pm$0.03 & 17.46$\pm$0.08 & 18.1$\pm$0.1 & 7$^{+92}_{-6}$ & 6.3$^{+1.2}_{-0.2}$ & 0.71 \\
3 & 36.275073 & -24.790066 & 16.00$\pm$0.01 & 15.97$\pm$0.01 & 16.42$\pm$0.03 & 17.02$\pm$0.04 & 54$^{+294}_{-50}$ & 7.6$^{+1.5}_{-1}$ & 1.04 \\
5 & 36.274900 & -24.786761 & 17.39$\pm$0.02 & 17.67$\pm$0.02 & 18.13$\pm$0.05 & 18.70$\pm$0.07 & 7$^{+291}_{-6}$ & 6.0$^{+2.3}_{-0.1}$ & 0.43 \\
6 & 36.274171 & -24.783779 & 17.17$\pm$0.01 & 17.25$\pm$0.02 & 17.65$\pm$0.04 & 18.19$\pm$0.07 & 54$^{+303}_{-50}$ & 7.1$^{+1.4}_{-1}$ & 0.6 \\
8-9 & 36.272113 & -24.784431 & 16.17$\pm$0.02 & 16.13$\pm$0.02 & 16.52$\pm$0.04 & 17.05$\pm$0.07 & 54$^{+50}_{-308}$ & 7.5$^{+1.4}_{-1.1}$ & 0.92 \\
11 & 36.266513 & -24.800236 & 16.63$\pm$0.02 & 16.96$\pm$0.02 & 17.45$\pm$0.04 & 18.07$\pm$0.06 & 7$^{+80}_{-6}$ & 6.3$^{+1.4}_{-0.4}$ & 0.78 \\
12-13 & 36.268845 & -24.801279 & 16.78$\pm$0.02 & 17.02$\pm$0.02 & 17.52$\pm$0.04 & 18.13$\pm$0.06 & 7$^{+279}_{-3}$ & 6.2$^{+2.3}_{-0.2}$ & 0.76 \\
14 & 36.270572 & -24.800231 & 16.77$\pm$0.01 & 17.02$\pm$0.01 & 17.53$\pm$0.03 & 18.14$\pm$0.05 & 7$^{+290}_{-6}$ & 6.2$^{+2.4}_{-0.3}$ & 0.71 \\
15 & 36.256402 & -24.797122 & 16.34$\pm$0.01 & 16.47$\pm$0.01 & 16.85$\pm$0.03 & 17.43$\pm$0.04 & 54$^{+268}_{-50}$ & 7.4$^{+1.4}_{-1.2}$ & 1.19 \\
16 & 36.264253 & -24.794724 & 16.26$\pm$0.005 & 16.05$\pm$0.008 & 16.47$\pm$0.02 & 17.05$\pm$0.02 & 54$^{+450}_{-50}$ & 7.5$^{+1.5}_{-1.3}$ & 1.19 \\
17 & 36.265672 & -24.793388 & 16.08$\pm$0.005 & 16.52$\pm$0.008 & 16.88$\pm$0.02 & 17.42$\pm$0.03 & 87$^{+456}_{-82}$ & 7.5$^{+1.4}_{-1.6}$ & 0.91 \\
18 & 36.262733 & -24.790521 & 17.51$\pm$0.005 & 17.36$\pm$0.007 & 17.72$\pm$0.02 & 18.28$\pm$0.02 & 54$^{+337}_{-50}$ & 7.0$^{+1.6}_{-1.4}$ & 0.67 \\
\enddata 
\tablecomments{The RA and DEC positions are in J2000 and approximations and are not necessarily related to the exact photometric center of the sources.\\}
\end{deluxetable*}

\subsection{Comparison with High Redshift Collisional Ring Clumps}
\label {goods}

The largest complexes in NGC 922 are close to the limit of resolved star-forming regions in galaxies in the Hubble Ultra Deep Field \citep[UDF][]{beck06} out to redshifts of about 4-5; W06 noted that several clump cluster galaxies in the UDF \citep{elm05} resemble NGC 922 in having outer rings. Galaxies in the GEMS and GOODS fields at redshifts 0.6 to 1.4 also include possible collisional ring galaxies \citep{elm06}. Two UDF galaxies, with catalog numbers 5190 and 9159, have large-scale star-formation in their outer ring-like structures \citep[see Fig.~1 in][]{elm05}. Their photometric redshifts are 1.25 and 1.45, respectively. They each have a diameter of about 10 kpc, and the complexes range from 0.5 to 2 kpc.  The largest complex is about 13\% of the galaxy size in UDF5190 and 20\% in UDF9159,  which is 3 times larger than the average for nearby late-type spirals but about the same as the ratio for NGC922, 15\% based on its diameter of 13.4 kpc (W06).  
The absolute complex sizes are about the same in the distant galaxies as in NGC 922, but the distant galaxies are physically smaller.
Photometric data for the clumps in these two galaxies were compared with evolutionary models that included bandshifting effects, interstellar reddening, and intergalactic H\,{\sc{i}} absorption \citep[see details in][]{elm05,elm06}. The clump ages are about $10^8$\,yr, and their masses range from 2$\times10^7$ to $2\times$10$^8$\,\msun. Thus, they are fairly similar to the largest complexes in NGC 922.  

A GOODS galaxy with a well-defined outer ring of star formation and an off-center nucleus is shown in Figure \ref{figMAST} (left) in color from the MAST archives\footnote{http:/archive.stsci.edu}. It is number 21238 in the COMBO-17 catalog, with a redshift of 0.664 \citep{wolf03}. Star-forming complexes are identified in Figure~\ref{figMAST} (right). Number 1 is the nucleus; 2 is the brightest star-forming region on the ring, and other ring clumps are 3-13. Features 14-16 are ``feathers''  on the western side, reminiscent of the plume in NGC 922, and 17 is a blue clump. Photometry was done on the features in the BVIz images.  Figure \ref{figcmd}  shows their color-magnitude diagram of V$_{606}$ versus (V$_{606}$-z$_{850}$). Overlaid are evolutionary tracks for clusters with different masses, as labeled. The open circles along the lines are timesteps, with two times indicated. Most of the clumps are 3-7$\times$10$^7$\,\msun. The feathers are in the same range. The blue dot, Feature 17, is much less massive, closer to 1$\times$10$^7$\,\msun. The nucleus is about 5$\times$10$^8$\,\msun. The ages of the ring clumps are 80 to 150\,Myr. The clumps on the southeast side of the ring are systematically redder than those on the northwest, possibly due to some reddening. 

%% FIGURE 12 %%
\begin{figure*}
\epsscale{1.0} 
%\plotone{GDS12-21238_cmyk_id.eps}
\plotone{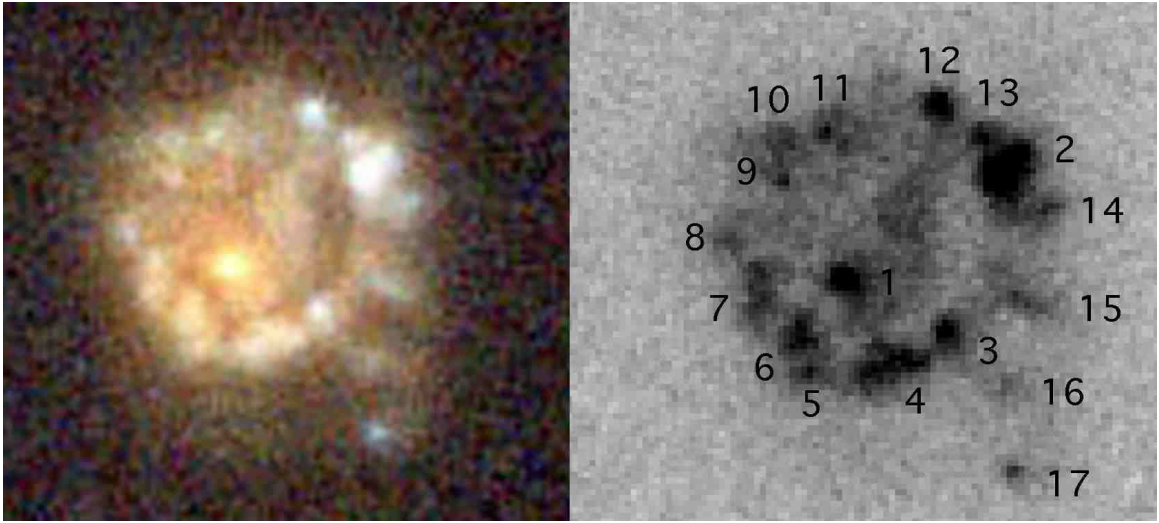}
\caption{Left: Collisional ring galaxy COMBO-17 21238 from the Hubble GOODS field number 12. Right: The complexes and features are labeled. See section~\ref{goods} for a detailed identification of the morphological features. The image is 3$\arcsec$ in length, or about 25\,kpc.\\}\label{figMAST}
\end{figure*}

Other GEMS and GOODS ring galaxies at redshifts z=0.7 to 1.4 have ring clumps that are about 0.1 times the galaxy diameter, with ages of a few$\times10^8$\,yr. Their masses range from several$\times10^6$\,\msun\ for the lower redshifts to a few$\times10^8$\,\msun for the higher redshifts \citep{elm06}.  These masses are slightly smaller than the giant clumps found in several non-ring clump cluster galaxies at redshift z=1.8 to 3.0, whose masses average 6$\times10^8$ M$_{\odot}$ \citep{elm05}.

\section{The Tidal Plume and Diffuse Emission of NGC\,922}
\label{tail}

The three-color image of NGC\,922 (Fig.~\ref{cmyk}) shows a faint yellow-red tail on the western side of the nucleus, pointing roughly toward the companion S2. This structure appears clearly in the WF2 chip. A smooth and unresolved stellar population comprises the tail; any clusters present are below the detection limit in U and B.

As was done for the nucleus and the two large clusters and nucleus (Section~\ref{data}), we made photometric measurements for the tail (see Table~\ref{complex}) in the area shown by a polygon in Figure~\ref{morph}. There was no detection in the F300W image. The photometric uncertainties depend mainly on the size of the polygon. 

Our population synthesis of the plume indicates stellar ages of 1 to 5\,Gyr, and a total stellar mass of the order of 7$\times$10$^8$\,\msun. The age found for the tail strongly suggests either that it was part of the main disk of NGC\,922, and was pulled away by S2 after the collision or it was part of S2 and was ripped off during the encounter. Based on the I-band residual image, we can see the tail extending up to 7-8\,kpc from star forming regions in the disk. If the stars come from NGC\,922's disk, this implies that the most distant stars have a projected linear velocity of about 20-30\,km\,sec$^{-1}$. Even though the material in the tail was subject to intense mixing and compressions, it did not produce significant star formation. This suggests that the plume may have formed from an ISM-free portion of the galaxy (i.e. bulge or thick disk).

%% FIGURE 13 %%
\begin{figure*}
\epsscale{0.7} 
%\plotone{v_v-z_GDS_rapid_decay1e7.eps}
\plotone{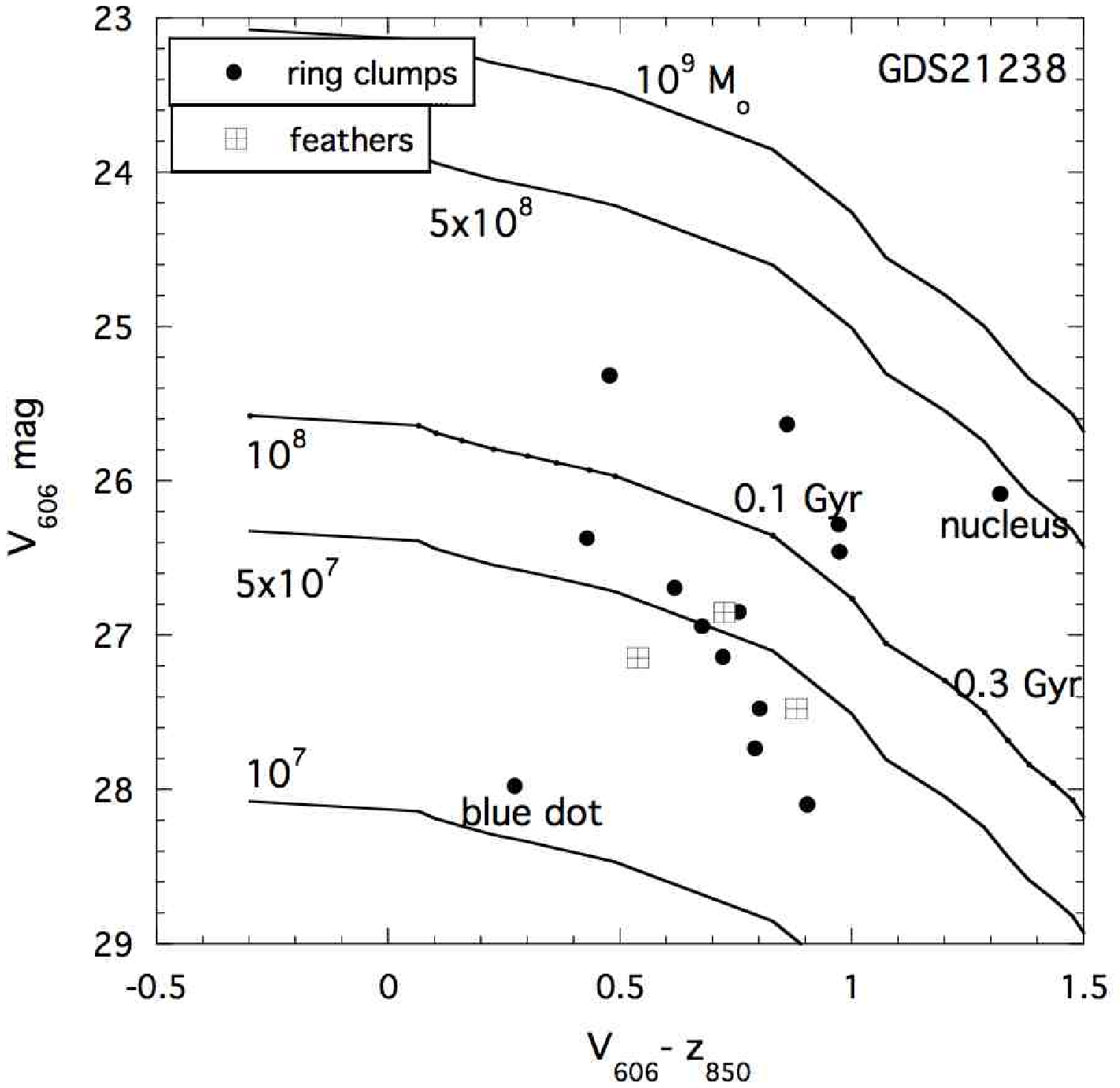}
\caption{Color-magnitude diagram showing the complexes and feathers in COMBO-17 21238, with evolutionary tracks overlaid showing cluster ages and masses.}\label{figcmd}
\end{figure*}

In addition to the tidal plume, another diffuse emission component is observed in the whole disk of NGC\,922 after removing the clusters from the images. We calculated the fraction of diffuse light and found that 79.2, 89.5, 93.0, and 94.1\% of the light was produced by this diffuse component for the U, B, V, and I filters, respectively. This indicates the fraction of light in clusters is more important at shorter wavelengths. The fraction values are also consistent with the work of \citet{meurer95} where they found that about 80\% of the UV light is found outside of star clusters in nearby starburst galaxies. The diffuse fraction is known to be higher in normal galaxies \citep{larsen00}. This indicates that NGC\,922 is more a starburst type of galaxy, i.e. with enhanced cluster formation activity. Synthesis modeling suggests an age of 250$\pm$40\,Myr for the diffuse emission, indicating that the diffuse stellar component was mainly formed before the collision $\sim$150\,Myr ago. We obtained that log(M$_{Stellar})=10.0\pm0.1$\,\msun\ for the diffuse emission, although the use of an instantaneous burst model might be less appropriate in this case.

\section{The Fate of NGC\,922 and its Star Clusters}
\label{fate}

According to general dynamical models, the ring in NGC\,922 will eventually be fragmented into several massive clumps, which would resemble high-redshift complexes and substructures. Its galactic disk will become thicker and it will evolve in morphology to become an S0. 

On a smaller scale point-of-view, \citet{bur05} proposed that collisional ring galaxies were excellent candidates for producing faint fuzzy clusters. They are known to be old, relatively large and non-compact clusters \citep[8-15\,Gyr, 7-15\,pc][]{lar00,bro02}. \citet{elm08} showed that they are actually the lucky survivors of star formation in regions where the gas density was barely enough to produce bound clusters but where the environment also presented low tidal forces so the clusters were not torn apart. The author explains that such environments can exist in the molecular clouds of outer galactic regions that are highly shocked such as in the expanding collisional rings of star formation. Consequently, the most massive older clusters (e.g. clusters $>$50\,Myr in Fig.~\ref{lumage}) observed in NGC\,922 are excellent candidates for progenitors of faint fuzzy clusters.

\section{Conclusions}
\label{conclusion}

In this work, we presented the four-band HST/WFPC2 images of NGC\,922, one of the nearest known collisional ring galaxies. We performed PSF photometry and detected 3817 stellar clusters in at least three bands, as well as two other clusters bigger than the PSF. We used the stellar population synthesis model Starburst99 to estimate ages and masses for each observed clusters. We also simulated the dynamical and star-forming events related to the collision between NGC\,922 and its companion S2. 

We found that most observed clusters are 7\,Myr old, or younger. Those clusters are mainly located within the ring or the bar, consistent with H$\alpha$ emission and in agreement with the simulation. A large fraction of clusters older than 50\,Myr are found in the nuclear region, which is consistent with the model producing many older clusters (100-200\,Myr) in the bulge. Those also tend to be massive, according to single stellar population modeling, but are probably contaminated by an older underlying stellar population which would then bias the age and mass estimation. 

We also observed a stellar tidal plume pointing in the direction of the companion. We did not detect clusters in the plume. We performed integrated photometry for this plume and found that it is consistent with stellar populations of 1 to 5\,Gyr old. This suggests that the material was stripped off from NGC\,922 by S2 and did not produce significant star formation related to the collision.

The cluster luminosity function observed for NGC\,922 has a slope typical of young cluster populations. The age distribution of star clusters shows a decreasing slope, consistent with cluster dissolution and disruption. By comparing the N-body/SPH star formation simulation to the observed age distribution corrected for cluster disruption (MI and MD models), we find that the MI model is over-correcting for dissolution while the MD model is under-correcting in the 10-100\,Myr age range. In any case, cluster dissolution and disruption is a plausible physical process to consider in order to reach good agreement between the simulation and observation.

We found a significant number of massive ($>$10$^5$\,\msun) star clusters in NGC\,922 that are old enough ($>$50\,Myr) to have survived the supernova explosion phase. Since some of them are likely to be born in highly shocked clouds and that they are now located at high galactic radii where the tidal disruption forces are presumably low, those clusters are excellent candidate progenitors of faint fuzzy clusters.

NGC\,922 has a morphology similar to several distant galaxies observed in the GEMS and GOODS fields. We detected about a dozen big star-forming complexes in the ring of NGC\,922.  A direct comparison with the knots observed in a ring galaxy from the GOODS field (\#21238 in COMBO-17, z$=$0.7) and other high redshift ring galaxies shows that NGC\,922's complexes are of comparable sizes and masses even though the distant galaxies are physically smaller.

\acknowledgments

AP is grateful to Barry Rothberg for support with HST/WFPC2 data and to R. Chandar for helpful scientific comments. This work was supported by the HST grant HST-GO-11112. GRM was also supported by NASA LTSA grant NAG5-13083. KB acknowledges financial support from the Australian Research Council (ARC). The numerical simulations reported here were carried out on SGI supercomputers at Australian Partnership for Advanced Computing (APAC) in Australia. 

\bibliography{mybib}
%\begin{thebibliography}{}
%
%\bibitem[{{Wong et al.}(2006)}]{wong06}
%{Wong}, O.~I. et al. 2006
%\end{thebibliography}

\end{document}